\documentclass[10pt,aps,prd, floats, floatfix, superscriptaddress, twocolumn,nofootinbib,amsmath,amssymb]{revtex4-2}

 \usepackage{enumerate}
 \usepackage{graphicx}
 \usepackage{multirow}
 \usepackage{xcolor}
 \usepackage{tablefootnote}
\usepackage{threeparttable}
 \usepackage{bm}
 \usepackage[normalem]{ulem}
 \usepackage{comment}

\usepackage{amsmath}
\usepackage{enumerate}
\usepackage{amssymb}
\usepackage{graphicx}
\usepackage{multirow}
\usepackage{xcolor}
\usepackage{tablefootnote}
\usepackage{threeparttable}
\usepackage{bm}
\usepackage[normalem]{ulem}
\usepackage{physics}
\usepackage{cancel}
\usepackage{soul}
\usepackage{tikz}
\usepackage{braket}
\usepackage{simpler-wick}
\usepackage{amsmath}
\usetikzlibrary{decorations.pathreplacing}

\usepackage{amsmath,mathtools}
\usepackage{float}

\usepackage[acronym]{glossaries}
 \usepackage{physics}

 {\left\lbrace\begin{array}{@{}l@{}}}%
 {\end{array}\right.}

 \newacronym{frw}{FRW}{Friedmann-Robertson-Walker}
 \newacronym{cmb}{CMB}{Cosmic Microwave Background}

\begin{document}

\title{A Bell Experiment in an Entangled Universe}

\author{Pablo Tejerina-P\'erez}
\affiliation{ICC, University of Barcelona, Mart\' i i Franqu\` es, 1, E08028
Barcelona, Spain}

\author{Daniele Bertacca}
\affiliation{Dipartimento di Fisica e Astronomia Galileo Galilei, \\Universit\`a degli Studi di Padova, via Marzolo 8, I-35131, Padova, Italy}
\affiliation{INFN, Sezione di Padova, via Marzolo 8, I-35131, Padova, Italy}
\affiliation{INAF- Osservatorio Astronomico di Padova, \\ Vicolo dell Osservatorio 5, I-35122 Padova, Italy}

\author{Raul Jimenez}
\affiliation{ICC, University of Barcelona, Mart\' i i Franqu\` es, 1, E08028
Barcelona, Spain}
\affiliation{ICREA, Pg. Lluis Companys 23, Barcelona, 08010, Spain.} 

\author{Leonid Sarieddine}
\affiliation{ICC, University of Barcelona, Mart\' i i Franqu\` es, 1, E08028
Barcelona, Spain}

\begin{abstract}
    We propose a possible quantum signature of the early Universe that could lead to observational imprints of the quantum nature of the inflationary period. Graviton production from the presence of a classical, coherent state of the inflaton scalar field results in entangled states in the gravitons' polarizations. At horizon crossing, interactions between the gravitons and (lower scale) inflatons, together with the gathering of ``which-path information'' from the cosmological horizon, perform the required Bell experiments leading to a definitive measure, which can be imprinted in the scalar correlation four-point function. This is because of a non-trivial effect due to the derivatives on two scalar fluctuations, and it provides a fingerprint that depends on the polarization of the graviton that Alice and/or Bob measured in their patch.  We hint how this signature could be measured in the high-order correlation function of galaxies, in particular on the halo bias and the intrinsic alignment.
\end{abstract}

\maketitle

\section{Introduction}
Inflation is a period of quasi-exponential expansion of the early Universe, i.e. a de Sitter stage with slightly broken time-translational symmetry to allow for a graceful exit \cite{Mukhanov:1981xt, Guth:1982ec, Hawking:1982cz, Bardeen:1983qw, Mukhanov_1}:
\begin{equation}
\label{eq: deSitter scale factor}
    a(t)\sim a_i\,e^{H_\Lambda t}\,\,,
\end{equation}
where $a$ is the scale factor, with initial value $a_i$ and $H_\Lambda = (8\pi G\varepsilon_\Lambda/3)^{1/2}$, where $\varepsilon_\Lambda$ is the constant energy density of a perfect fluid with positive cosmological constant $\Lambda$, with equation of state $p_\Lambda = -\varepsilon_\Lambda$.

The presence of an inflationary stage previous to the Big Bang (or what used to be called ``Big Bang'' before the upcoming theory of inflation) solves in a natural manner the fine tuning and naturalness problems of the previous theory (horizon problem, flatness problem/fine tuning of the initial conditions, and others - for reference, see for example \cite{Mukhanov_1}) while keeping its successful predictions like Big Bang Nucleosynthesis. It also introduces a natural way to explain the inhomogeneities and anisotropies in the Universe that gave rise to the formation of the cosmological structures we observe today. Quantum fluctuations in the fields present during inflation (in its most minimal form, just the spacetime-metric field, and a scalar field called the \textit{inflaton}) get stretched by the exponential expansion to cosmological distances. Their ``size'' (physical wavelength) grows proportional to $a$, while their amplitude decays as $a^{-1}$. The curvature scale $H_\Lambda^{-1}$ (or Hubble horizon) remains almost constant during inflation. Thus, for any mode of fixed comoving wavenumber $k$, the physical wavelength $\lambda_{ph}=2\pi(a/k)$ of a quantum fluctuation generated inside the Hubble horizon will soon become larger than $H_\Lambda^{-1}$. Therefore, the corresponding mode leaves the horizon, and starts ``feeling'' the curvature of spacetime. At horizon crossing, its amplitude freezes and remains almost constant until the end of the inflationary stage. Once inflation is over, the radiation dominated era starts (decelerating Hubble expansion), during which the curvature scale $H_\text{rad}^{-1}$ starts growing at a rate $a/\Dot{a}$. At this point, the now large fluctuations reenter the Hubble horizon as density perturbations of cosmological scale, and as gravitational waves. These density perturbations work as classical gravitational seeds for the formation of large-scale structures. Modes that left the horizon the latest will be the first ones to reenter, and vice versa. For the modes of cosmological interest, i.e. those responsible for the observed large-scale structure of the Universe (between $1$ Mpc and $3000$ Mpc approximately), the moment of horizon crossing should be early enough\footnote{These modes left the horizon around $8$ e-folds after the beginning of inflation, assuming it lasted for $60$ e-folds (around the minimum necessary to solve the horizon problem) \cite{Kolv_Turner}.}, so that they re-enter around the present observable large scale today.

In this context, there is a somewhat natural question that emerges: at what point does a quantum fluctuation generated during inflation stop being of quantum nature? The analysis that we apply to the Cosmic Microwave Background (CMB), which provides us with the earliest observational features of the Universe so far, and to large scale structure survey's data, is classical. But where did the quantum nature of the gravitational seeds go? It is still an open question how this classicalization of quantum fluctuations occurs. If any signal of the early nature of the quantum fluctuation remains, this should be imprinted in some current observable, maybe in the form of non-gaussianities in the CMB anisotropies, or in higher-order correlation functions of galaxy distributions, or some other observable. 

There has been previous work in the literature trying to search for quantum signals of the early Universe\cite{Campo:2003pa,Campo:2003gb,Campo:2005sv,Campo:2005qn,Campo:2007ar,Campo:2008ju,Campo:2008ij,Choudhury:2016cso,Choudhury:2016pfr,Kanno:2015ewa,Martin:2015qta,Kanno:2017teu,Martin:2017zxs,Martin:2018zbe,Martin:2018lin,Kanno_2019,kanno2020polarized,Danielson:2021egj,Colas:2022kfu,Prabhu:2022zcr,DaddiHammou:2022itk,Ning_2023}. Quantum discord (a measure of ``\textit{quantumness}'') of inflationary perturbations is calculated in \cite{J.Martin_V.Venin} and suggests some features to probe different levels of discordance in CMB descriptions. Possible observational signatures in the CMB of graviton exchange between tensor and scalar fluctuations are discussed in \cite{Bellomo_2018}. On the other hand, due to the environment, the potential decoherence of quantumness upon classicalization has been widely  discussed in literature, e.g., see \cite{Calzetta:1995ys, Lombardo:2005iz, Martineau:2006ki, Kiefer:2006je, Kiefer:2008ku, Nelson:2016kjm, Burgess:2014eoa, Martin:2018zbe, Boyanovsky:2015tba, DaddiHammou:2022itk, Burgess_decoherence, Sou:2022nsd, Ning_2023}. 
For example, in \cite{Collins_2016}, they discuss how the CMB polarization components $\bf E$ and $\bf B$ are modified due to entanglement of scalar and tensor fluctuations. Regarding primordial gravitons, a possible source of decoherence should also include the nonlinear interaction between tensor modes \cite{Gong:2019yyz} and the scalar-tensor interaction \cite{DaddiHammou:2022itk, Burgess_decoherence, Sou:2022nsd}. Another possibility is to add extra fields to the simplest model of inflation such that one can construct Bell inequalities as done in \cite{Maldacena_2015}. 
Finally, on modifications to Cosmological Power Spectra due to
entanglement effects in the initial quantum state of scalar-tensor fluctuations, e.g., see \cite{Bolis:2016vas}.

In section \ref{sec: Entanglement and bell ineq}, we introduce the concept of entanglement and Bell inequalities. In section \ref{sec: Inflation as our lab}, we suggest a mechanism by which entangled states are created during inflation, via the interaction of gravitons and an inflaton in a classical, coherent state, such that it acts as a ``pump'' field. Then, we describe plausible processes by which the quantum nature of the tensor fluctuations of the metric field (i.e. gravitons) during inflation is made explicit. We discuss how, through interaction with their environment, gravitons may imprint this quantumness into some observable quantity. We then propose what this observable quantity might be. In an appendix~\ref{sec: Perturbations in inflation}, we review the context in which perturbations get generated during inflation, and introduce the relevant hamiltonian for the creation of an entangled state of gravitons.

\section{Quantum nature, entanglement, and Bell inequalities}
\label{sec: Entanglement and bell ineq}

In this section we briefly recall the essential elements of quantum entanglement and Bell inequalities needed for our inflationary construction. The purpose is not pedagogical completeness, but to fix notation and clarify the logical ingredients required to identify genuinely quantum correlations in cosmological observables.

\subsection{Entanglement and non-separability}

Entanglement is present when the quantum state of a composite system cannot be written as a product of states associated with its subsystems, i.e. when the state is \emph{non-separable}. For a bipartite system with basis states $\ket{\psi_1}$ and $\ket{\psi_2}$, a generic entangled state can be written as
\begin{equation}
\label{eq: generally entangled 2 state (no prescription)}    
    \ket{\Psi}
    =\alpha\,\ket{\psi_1}_1 \otimes \ket{\psi_1}_2
    +\beta\,\ket{\psi_2}_1 \otimes \ket{\psi_2}_2 \, ,
\end{equation}
with $\alpha,\beta\in\mathbb{C}$ and $|\alpha|^2+|\beta|^2=1$. Here $\ket{\psi_i}_j$ denotes subsystem $j$ prepared in state $\ket{\psi_i}$.

Entanglement arises solely from the principle of quantum superposition and implies correlations that cannot be reproduced by local classical descriptions. Upon measurement of an observable $\mathcal{O}$ on subsystem~1, the global state $\ket{\Psi}$ collapses to a definite branch, thereby fixing the state of subsystem~2 instantaneously, irrespective of their spatial separation. While each subsystem is individually described by a mixed state, the full bipartite system remains in a pure quantum state. These nonlocal correlations persist even when the subsystems are causally disconnected.

\subsection{Bell inequalities and experimental structure}
\label{subsec: Bell ineq and experiment}

Bell’s theorem \cite{Bell} provides a sharp distinction between correlations allowed by local hidden-variable theories and those permitted by quantum mechanics. For our purposes, we emphasize the minimal ingredients required to construct a Bell-type experiment, following \cite{Maldacena_2015}:

\begin{itemize}
    \item Two spatially separated regions, denoted Alice and Bob.
    \item An entangled bipartite state of the form \eqref{eq: generally entangled 2 state (no prescription)}, with one component in each region.
    \item Local measurements of a given observable, parametrized by variables $\theta$ and $\theta'$ for A, and $\phi$ and $\phi'$ for B. The observable comes from the expectation value of non-commuting operators $A(\theta), A(\theta')$ for Alice and $B(\phi), B(\phi')$ for Bob.
    \item Binary measurement outcomes, normalized to $\pm1$.
    \item A classical channel allowing the comparison of outcomes.
\end{itemize}

One defines the CHSH combination
\begin{equation}
\label{eq: S observable}
    S = C(\theta, \phi) + C(\theta', \phi)
      + C(\theta, \phi') - C(\theta', \phi') \, ,
\end{equation}
where $C(\theta, \phi)=\expval{A(\theta)\,B(\phi)}$. This quantity is defined as an average over many realizations of the same 2-state. Local hidden-variable theories satisfy
\begin{equation}
    |S| \leq 2 \, ,
\end{equation}
while quantum mechanics allows
\begin{equation}
    |S| \leq 2\sqrt{2} \, ,
\end{equation}
the latter being achieved for maximally entangled states \cite{CHSH}. An observed violation of the classical bound therefore constitutes direct evidence of intrinsically quantum, nonlocal correlations.

This framework will be mapped in Sec.~III onto an inflationary setting, where the role of the entangled subsystems is played by graviton polarizations, and the measurement and classical communication are implemented through local interactions and horizon-induced decoherence.

\begin{figure*}
    \centering
    \includegraphics[trim=40 40 50 40, clip, width=0.7\textwidth]{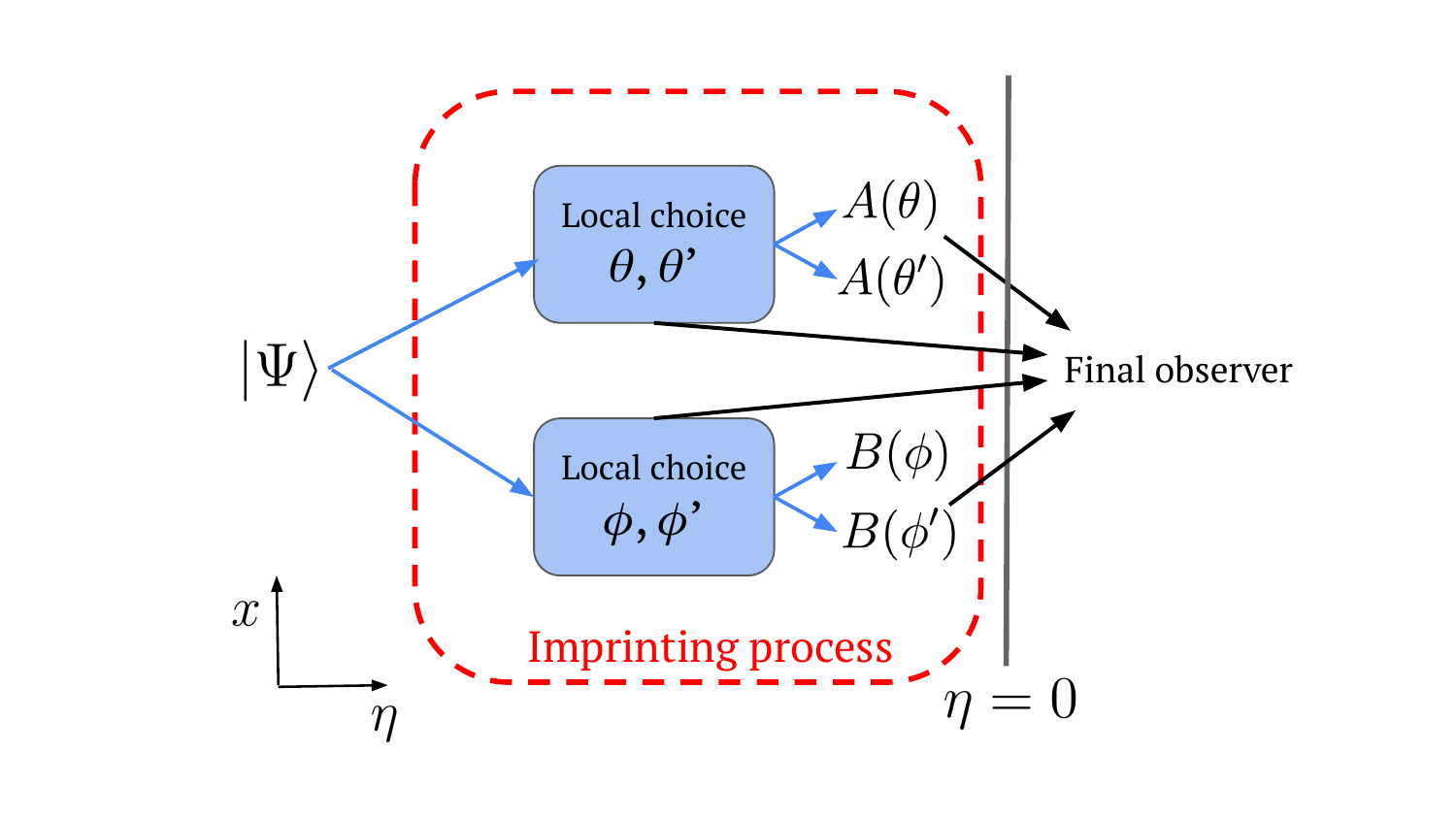}
    \caption{Set up for a Bell inequality violating experiment. The comoving time flows from left to right ($\eta=0$ represents the end of inflation), and spacial dimensions are drawn vertically. The entangled 2-state $\ket{\Psi}$ has components at two spatially-separate locations 1 (top) and 2 (bottom). Two different results of the measurement of observables $A$ and $B$ can be obtained, each dependent on the choice of the local variables $\theta_1$ and $\theta_2$, respectively. This choice is dependent on the interaction of the state with its local environment, it can be seen as a dependence on the measuring apparatus. The possible imprinting of the entanglement of state $\ket{\Psi}$ is represented by the dashed-line box. These results, together with the choices of $\theta_i$, are transmitted through classical channels (black arrows) to a final observer after inflation has finished.}
    \label{fig: maldacenas bell experiment setup}
\end{figure*}

\subsection{Bell states}

As we have seen, an entangled 2-state is generally of the form \eqref{eq: generally entangled 2 state (no prescription)}. It is useful to work with 2-states known as \textit{Bell states} \cite{Gerry}, which are defined as
\begin{align}
    & \ket{\Phi\text{Bell}^\pm}=\frac{1}{\sqrt{2}}(\ket{0}_1\ket{1}_2\pm \ket{1}_1\ket{0}_2)\,, \label{eq: bell state psi +-} \\
    & \ket{\Psi\text{Bell}^\pm}=\frac{1}{\sqrt{2}}(\ket{0}_1\ket{0}_2\pm \ket{1}_1\ket{1}_2)\,. \label{eq: bell state phi +-}
\end{align}
Here, $\ket{0}_i$ and $\ket{1}_i$ represent the basis states of a generic bipartite system\footnote{A bipartite or two-mode system is a single-particle system with only two non-degenerate eigenvalues for a given observable $\mathcal{O}$, such that the one-state can be written as $\ket{\psi}_i^\text{1-state}=a\ket{0}_i + b\ket{1}_i\,\,$, with $a,b\in\mathbb{C}$, and $|a|^2 + |b|^2 = 1$.} (or two-mode system). The notation $\ket{i}_1\otimes \ket{j}_2$ denotes the tensor product of the basis states of subsystems 1 and 2, which define a basis of the Hilbert space $\mathcal{H}=\mathcal{H}_1\otimes\mathcal{H}_2$, in which two-particle states live. \\
As described extensively in the literature (see for example \cite{CHSH, Gerry}), the value of the observable $S$ in Eq. \ref{eq: S observable} can be larger than $2$ for these Bell states (thus rendering a result non accessible through local hidden variables theories). It is useful then to work with Bell states when aiming for a violation of Bell inequalities. Since one of the aims of this work is to design a cosmological Bell experiment, we will look for a mechanism that yields a Bell state of the pertinent quantum fields in our cosmological setup (see section \ref{sec: Inflation as our lab}).

\section{Inflation as our laboratory} \label{sec: Inflation as our lab}

We have described all the necessary ingredients to design a plausible cosmological Bell-type experiment. We now propose how to obtain all these ingredients in our inflationary setup. Some signature of violation of Bell inequalities, and therefore of the non-locality intrinsic to the quantum theory, would allow us to prove the quantum mechanical origin of primordial in-homogeneities that gave rise to the observed large scale structure of our Universe. At least, this possible signature might serve as prove of concept for the theoretical scientific community, and serve as a possible observational path to follow by the experimental community.\\

A review of scalar and tensor perturbations is given in appendix~\ref{sec: Perturbations in inflation}. Although this is textbook material, it helps as a guide for nonspecialized readers making the article self-consistent and also to set the notation in the article.\\

\subsection{Elements of the cosmological Bell experiment} \label{subsec: elements of Bell exp}

We proceed to describe our realization of a cosmological Bell-type experiment in the inflationary epoch. For that we need a specific realization of the general elements described in section \ref{subsec: Bell ineq and experiment} in our inflationary setup, and goes as follows:

\begin{itemize}

    \item Alice's and Bob's locations will be two spatially separate halos at the inflationary Hubble horizon, $H^{-1}_\Lambda\approx \text{constant}$. The distance $L$ between halos A and B will be set by the inverse of the momenta of the perturbations used in the Bell experiment $L\sim 1/k$ (see points below). The polarization of the tensor perturbations will be recorded when their momenta is of order of the horizon.

    \item Our Bell state describes two gravitons entangled in their tensor polarizations, see section \ref{subsec: entangled state of gravitons}, Eq. \eqref{eq: entangled state of gravitons}.

    \item The physical observable is a correlation function coming from tensor-scalar-scalar interactions, that depend on the polarization of the gravitons, with two possible states $P_i = +, \times$ for $i=1,2$ (one could use also L and R polarizations by changing the imprinting procedure). The measure at each location is performed by a series of steps involving these scalar-tensor interactions and a decoherence effect due to the separation into system-environment parts. The explicit steps are given in section \ref{subsec: process - general idea}.

    \item The possible results of the measurement (decoherence) of Alice's and Bob's gravitons are values $\pm 2$ depending if the state of the graviton after decoherence/measure is $+$ or $\times$ polarization. In our case, the result would then leave an imprint through the interaction of four inflatons with the polarized background curvature (generated by the exchange of a graviton). We denote this process as graviton exchange (GE), described in section \ref{subsec: Imprint through GE} (see also \cite{Bellomo_2018} or \cite{Seery_2009_Trispec_GE} for a discussion on GE). Intuitively, one could say that both Alice and Bob use a pair of scalar fluctuations each to imprint the results of their measure of polarization of the graviton. Note that we have a non-trivial effect due to the derivatives on the two scalar fluctuations, which translates into a non-trivial fingerprint that depends on the polarization of the graviton\footnote{This argument is better seen in the action \eqref{eq: action for GE} and the proportionality factor of GE in Eq. \eqref{eq: sum over polariz}, or also in Eq. 5.1 in \cite{Burgess_decoherence} or 3.15 in \cite{Maldacena:2002}. One can see that the presence of derivatives in the scalar field that appear in the relevant action for GE, i.e. Eq \eqref{eq: action for GE}, implies (once taken to Fourier space) a contraction of the scalar momenta with graviton polarizations. These scalar fields are the ones that correspond to Alice's and Bob's measurements in their patch.}.

    \item The result of the measure (or its imprint) is classically transmitted by the propagation of these pairs of scalar fluctuations after horizon crossing, where their amplitude freezes until the end of inflation, and then reenters during radiation epoch. The results are transmitted to a common location, constrained by the range of momenta/scales corresponding to Alice's and Bob's scalar fluctuations at late-times. Results can then be correlated at the common location.
\end{itemize}

The derivatives on the scalar field fluctuations could leave an additional imprint on the two subhalos of each patch (see Fig. \ref{fig: Imprinting}), with the possibility that we could observe it on the large scale structure through the ``intrinsic alignment''. Note that this ``intrinsic alignment'' is obtained by performing the perturbed Taylor expansion between one subhalo with the second one in the same patch. In this case, the two derivatives are obtained for the same potential, leading a tidal effect between these two subhalos linked to the polarization of the graviton in that patch. Then, correlating with the intrinsic alignment between subhalos of the other patch, one could again probe entanglement between the original gravitons.\\

In the following, we discuss more closely each of these elements of our inflationary setup.

\subsection{Entangled state of gravitons} \label{subsec: entangled state of gravitons}

We take as our state of study for the Bell experiment in the inflationary setup a quantum state of 2 gravitons entangled in their polarizations, being the state:

\begin{align}
\label{eq: entangled state of gravitons}
    \ket{\Psi_\text{Bell}^-} & = \frac{1}{\sqrt{2}} \Bigl(\ket{+,\mathbf{p}_1}_1 \otimes \ket{+,\mathbf{p}_2}_2 - \ket{\times,\mathbf{p}_1}_1 \otimes \ket{\times,\mathbf{p}_2}_2\Bigr) \notag \\
    & \equiv \frac{1}{\sqrt{2}}\Bigl(
    \ket{+}_A\ket{+}_B - \ket{\times}_A\ket{\times}_B\Bigr)\,,
\end{align}
where in the second line we have rewritten the states as $\ket{s_i, \,\mathbf{k}_i}_1\otimes\ket{s_j, \,\mathbf{k}_j}_2\equiv\ket{s_i}_A\ket{s_j}_2$, having also associated the state of graviton 1 to Alice, and the state of graviton 2 to Bob. One can see this state is a Bell state of the form \eqref{eq: bell state phi +-}.\\

We propose two alternative ways that could give rise to such kind of entangled states for gravitons during inflation. One possibility would be a scattering process of two incoming inflatons scattering into two gravitons entangled in state \eqref{eq: entangled state of gravitons}. This scattering can be realized with two 3-vertex interactions from the perturbed action \eqref{eq: action 1} to third order action (combining scalar and tensor fluctuations), or from the single 4-vortex involving $\zeta\zeta\gamma\gamma$ from perturbing the same action to fourth order, giving the expression in \eqref{eq: 4th order zeta zeta gamma gamma action}. A more detailed calculation of the later example is given in appendix \ref{subappendix: 4th order 2-to-2 scattering}.\\

The second possibility we propose for the production of entangled gravitons is the decay of an scalar fluctuation of momentum $\mathbf{p}$ into a pair of gravitons of momenta $\mathbf{k}_1,\,\mathbf{k}_2$ through a 3-vertex involving $\zeta\gamma\gamma$. In our calculation (see appendix \ref{subappendix: inflaton decay calculation}), we find that entanglement in polarization arises when the momenta of the gravitons is approximately opposite, $\mathbf{k}_1 \approx -\mathbf{k}_2$. This implies $||\mathbf{p}||\ll~ ||\mathbf{k}_1||, ||\mathbf{k}_2||$, which in turn would imply that $m_\phi \sim~||\mathbf{k}_1||$. Since we want the gravitons to be produced well inside the cosmological horizon, we have $||\mathbf{k}_1||,||\mathbf{k}_2||\gg \mathcal{H}$. In the simplest scenario, and because we are in a slow-roll regime, we have the mass of the inflaton field 
$m^2_\phi \propto V_{,\,\phi\phi} \ll H^2$, which renders the gauge invariant scalar fluctuation effectively massless \cite{Baumann_book_2022, Mukhanov_2}, and directly suppresses the decay $\zeta\rightarrow \gamma\gamma$. To solve this problem, one can include in ones model the presence of an extra scalar field $\psi$ that couples to the inflaton in the desired manner to increase the mass for a short period of time. For example, we could hypothesize an equilibrium state of the extra field $\psi$ under some symmetry. Because of the expansion, this symmetry would spontaneously break at some early inflationary time $\bar{\eta}\ll\eta_\text{eq}$ (to keep $k>k_\text{eq}$) and introduce a coupling like $\sim m^2(\psi)\phi^2/2$, increasing the mass of the inflaton field $m^2_{\phi,\text{tot}} = m_\phi^2 + m^2(\psi) > H^2$. This increase in mass triggers the creation of entangled graviton pairs in the state \eqref{eq: entangled state of gravitons}. After a short period of time, the extra scalar field $\psi$ decouples from the inflaton field and our setup returns to a slow-roll regime of inflation.

\subsection{Measurement, decoherence, and imprinting of entangled graviton polarizations--the general idea}
\label{subsec: process - general idea}

In this subsection we formalize the measurement process that converts the entangled graviton state generated by the inflaton pump into correlated classical records in spatially separated regions. The logical structure of the process is summarized schematically in Fig.~2, while its microscopic realization is illustrated in Fig.~5.

\subsubsection{From entangled production to spatial separation}

As shown in Sec.~III.B, the interaction between a coherent inflaton background and the tensor sector produces a large number of graviton pairs in an entangled Bell state,
\begin{equation}
\ket{\Psi_\text{Bell}^-} \;\propto\; 
\ket{+}_A\ket{+}_B - \ket{\times}_A\ket{\times}_B \, ,
\end{equation}
where $+$ and $\times$ label the two graviton polarizations. The two gravitons propagate into distinct Hubble patches, denoted by $A$ (Alice) and $B$ (Bob), as depicted in Fig.~2. At this stage the state is fully quantum and no definite polarization has been selected in either patch.

\subsubsection{Local recording of polarization}
\label{subsec:recording}

The first step of the measurement occurs locally in each patch through interactions between the graviton and additional inflaton quanta. Concretely, the graviton participates in scattering processes mediated by the scalar--tensor interaction, as illustrated in Fig.~5. Through these interactions, the polarization degree of freedom of the graviton becomes entangled with local inflaton modes.

From the point of view of quantum measurement theory, this interaction constitutes a ``recording'' of the graviton polarization: the inflaton degrees of freedom act as a local pointer basis that becomes correlated with either the $+$ or $\times$ polarization. Importantly, this process is local to each patch and does not, by itself, break the entanglement between the two gravitons.

\subsubsection{Environment-induced decoherence and classicalization}
\label{subsec:decoherence}

The transition from a recorded quantum state to a classical outcome is driven by decoherence induced by the separation of the system into system-environment subsystems \cite{kiefer1998coherence, polarski1996semiclassicality, Burgess_decoherence}. Specifically, decoherence implies that the off-diagonal elements (coherence-terms) of the quantum density matrix undergo exponential damping terms. Upon horizon crossing of the graviton modes under consideration in our quantum state, defined by
\begin{equation}
k|\eta| \simeq 1 \, ,
\end{equation}
one obtains decoherence by tracing out the environment degrees of freedom corresponding to the modes still inside the horizon that interact with them \cite{Burgess_decoherence}. Then, the state of the gravitons can be understood as a mixed state, where each graviton can be in a $+$ or $\times$ polarization, and there is a classical statistical uncertainty on the selected polarization state. Importantly, since the initial state of the gravitons was the Bell-like state of Eq. \eqref{eq: entangled state of gravitons}, even after decoherence, the polarization state of both gravitons is forced to be the same, with a classical uncertainty on which one:

\begin{equation}
    \ket{\Psi_\text{Bell}^-}\overset{\text{Decoherence}}{\longrightarrow} \Bigl\{ \ket{+}_A\ket{+}_B \ \text{or} \ \ket{\times}_A\ket{\times}_B  \Bigr\}
\end{equation}
where the notation $\{\ \cdot\ \ \text{or} \ \ \cdot \ \}$ expresses the statistical uncertainty discussed above. The fact that they are both on the same polarization state will be crucial for our construction.

\subsubsection{Nonlocal correlation of outcomes}

As explained above, because the initial graviton state is entangled, decoherence in one patch enforces a correlated outcome in the other. If the graviton in patch $A$ decoheres into the $+$ polarization, then the graviton in patch $B$ decoheres into the same polarization, and likewise for $\times$. This correlation is not mediated by any causal interaction between the patches; it follows directly from the entangled structure of the state prior to decoherence.

As a result, both Alice’s and Bob’s patches contain consistent classical records of the same polarization outcome, encoded in their local inflaton degrees of freedom. These correlated records form the basis for the Bell-type correlations represented schematically in Fig.~2.

\subsubsection{Imprinting and transmission to late times}

The final step is the imprinting of the polarization-dependent outcome onto scalar observables. As shown in Fig.~\ref{fig: Imprinting}, GE between inflaton pairs induces polarization-dependent contributions to higher-order scalar correlation functions. Since these interactions occur before horizon crossing of the relevant scalar modes, the resulting imprints are frozen into the curvature perturbations and subsequently transmitted as classical information to late times. We present a specific construction of the imprinting mechanism in the next sub-section \ref{subsec: Imprint through GE}.

In this way, the sequence
\begin{align*}
\text{entangled production}
\;\rightarrow\;
\text{local recording}
\;\rightarrow\; \\
\text{environment-induced decoherence}
\;\rightarrow\;
\text{classical imprint} \nonumber
\end{align*}
provides a complete and self-consistent measurement process. 

This framework explains how a genuinely quantum phenomenon—the production of entangled gravitons during inflation—can give rise to classical, spatially separated, yet nonlocally correlated observables.

\subsection{Construction of the imprinting mechanism based on GE -- the explicit prescription} \label{subsec: Imprint through GE}

\begin{figure}
    \centering
    \includegraphics[width = 0.45\textwidth]{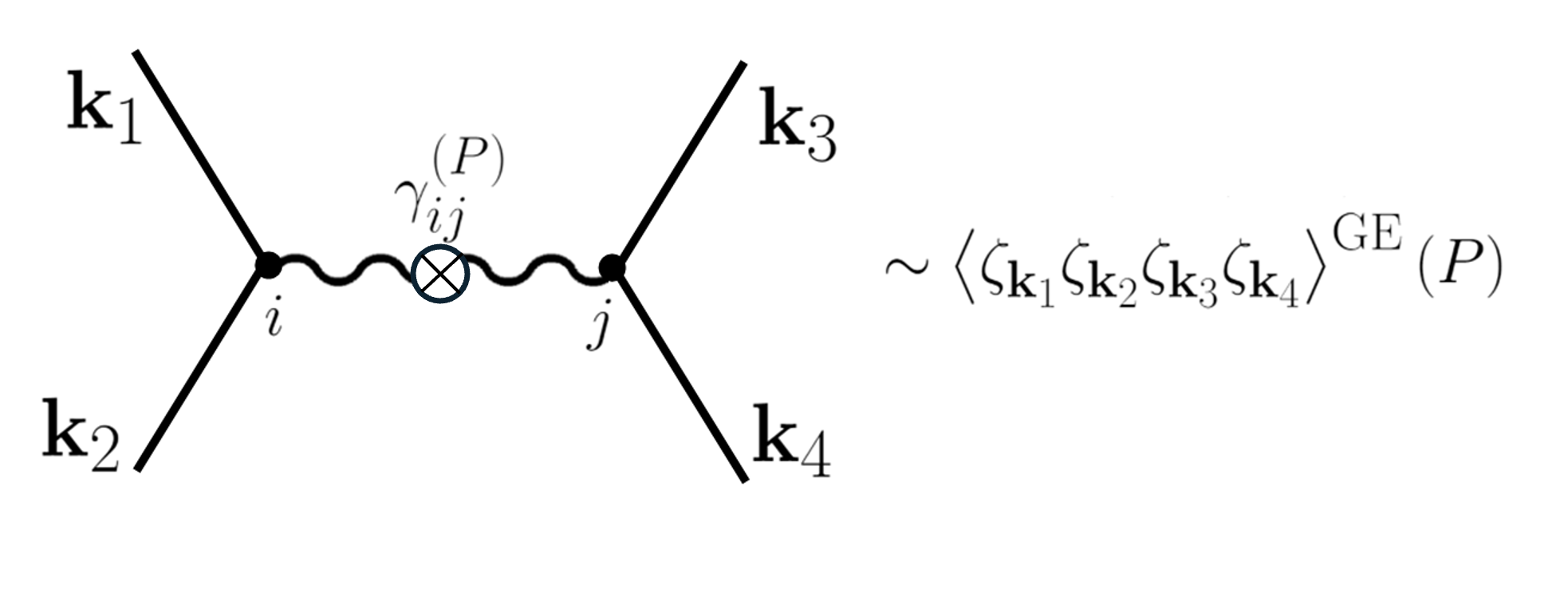}
    \caption{Graviton exchange of four inflatons \cite{Bellomo_2018}. Process of imprinting at horizon crossing. The graviton is entangled with its partner, as described in the text. Note that the graviton on the diagram is in fact the two soft gravitons. The dependence in $P$ stretches the fact that this process will depend on the polarization of the graviton, as described in the text.}
    \label{fig: feynmann diag for 4 scalar int}
\end{figure}

Let us now make a specific construction for the imprinting of the non-local correlations coming from the gravitons' entangled state.

First, we note that the polarization of the \textbf{graviton}, which has a discrete spectrum ($P=+,\times$), is the magnitude that \textbf{induces the non-local correlations} through \textbf{entanglement}. 

\begin{figure}
    \centering
    \includegraphics[width=0.49\textwidth]{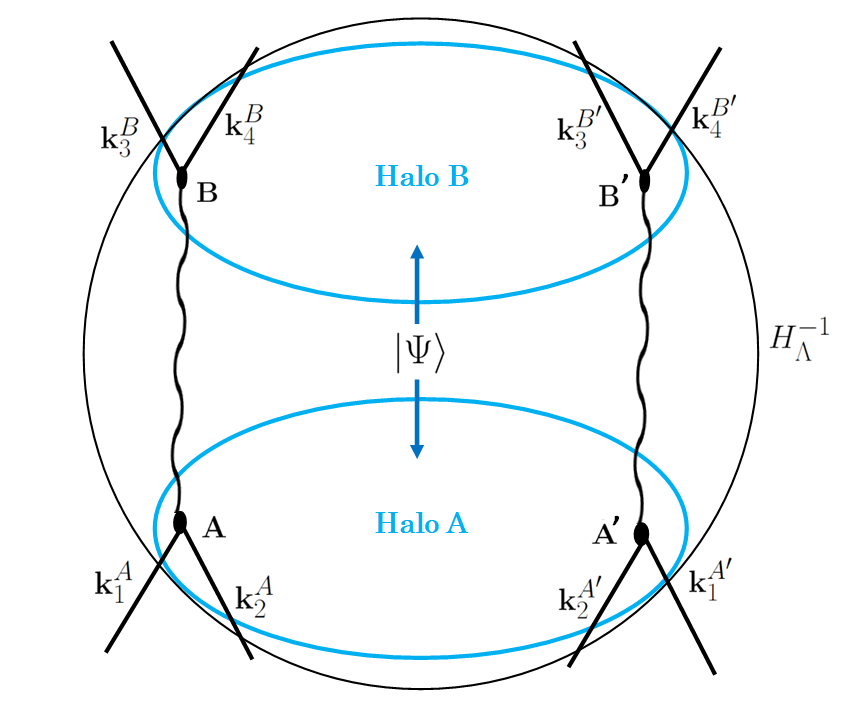}
    \includegraphics[width=0.49\textwidth]{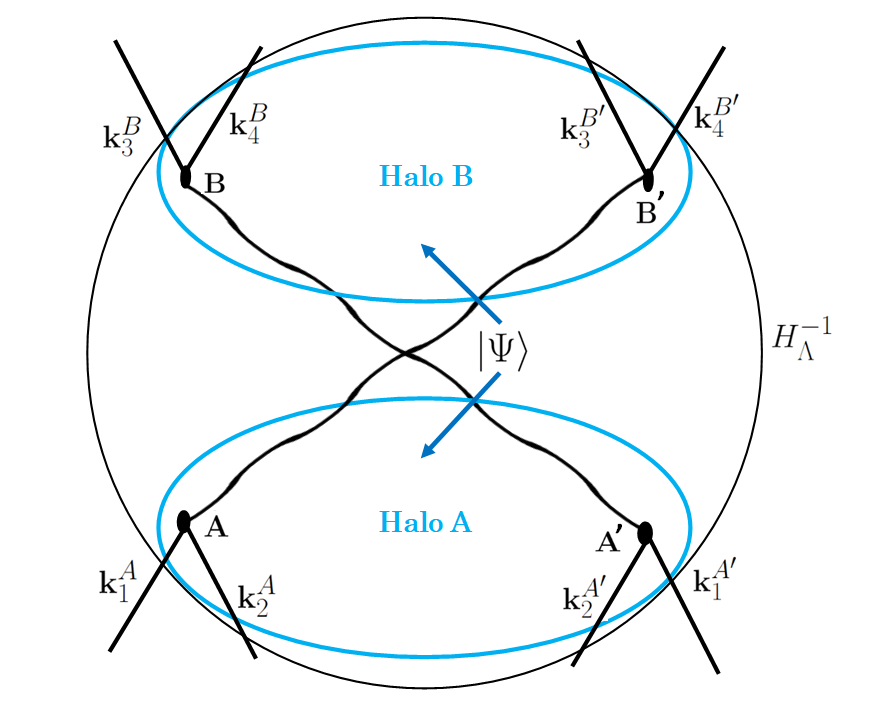}
    \caption{Imprinting of non-local correlations between halos $A$ and $B$. The left figure represents the first and last term of equation \eqref{eq: S for correlation functions} (terms $AB$ and $A'B'$), while the right figure represents the second and third terms ($A'B$ and $AB'$). The cosmological horizon is labeled as $H_\Lambda^{-1}$. The process is interpreted as follows. The 2-state wave function $\ket{\Psi}$ propagates to both halos, as gravitons propagate in approximately opposite directions towards halos A and B, respectively. Each wavy line represents the exchange of a low energy graviton between inflatons (straight lines) at different locations in each halo (subhalos without and with ``prime''). The GE occurs at a time before horizon crossing of the gravitons, which imprints the non-local effects of entanglement in polarizations. At a later time, the inflatons (whose correlation carries the imprinting) become super-horizon.}
    \label{fig: Imprinting}
\end{figure}

Now, we suggest that an \textbf{imprinting of the non-local correlations} occurs on pairs of \textbf{scalar} fluctuations that undergo GE. We suggest that the imprinted scalar pairs can the be used to construct some observable potentially measurable today, through their 4-point correlation function\footnote{As described in \cite{Bellomo_2018}, the graviton exchange contribution to the 4-point function of curvature perturbations (inflatons) leaves a specific signal, thus also affecting the clustering properties of collapsed structures. Concretely, in \cite{Bellomo_2018} it is shown that GE has a contribution to the two- and three- point function of dark matter halos - halo bias, or in the different directions in which the intrinsic alignment of galaxies is computed).}. To this end, we consider 2 pairs of (sub-horizon) inflatons in two separate halos/patches ($A$ and $B$), one pair per subhalo. This is, we consider one pair of inflatons with momenta $\mathbf{k}_1$,  $\mathbf{k}_2$ at subhalo $A$, one pair $\mathbf{k}_3$, $\mathbf{k}_4$ at subhalo $B$, one pair $\mathbf{k}_1'$, $\mathbf{k}_2'$ at subhalo $A'$, and one pair $\mathbf{k}_3'$, $\mathbf{k}_4'$ at subhalo $B'$ (see Fig. \ref{fig: Imprinting}). These scalar fluctuations will be the ones undergoing GE, as described in the following paragraphs.\\

Generally, GE is a contribution to the 4-point function of curvature perturbations, given two pairs of inflatons exchanging one graviton, as in Fig. \ref{fig: feynmann diag for 4 scalar int}. Its vertex is given by the action \cite{Seery_2009_Trispec_GE, Bellomo_2018}:
\begin{equation}
   I^{(3)} \supset \int \text{d}t\text{d}^3\mathbf{x}\,M_p^2 \epsilon_1 a \gamma^{ij}\partial_i\zeta \partial_j\zeta 
   \label{eq: action for GE}
\end{equation}

The reason we use GE is the way it couples the variable that carries the entanglement (polarization of the graviton) with the variable in which non-local correlations get imprinted (scalar perturbations). These scalar perturbations, after classicalization, are also the classical channel - together with expansion - of transmission of the measurements. 

Let us analyze how GE couples the polarization of the gravitons (which are in an entangled state) to the scalar perturbations \textit{before} the decoherence process, and leave an imprint on their 4-point correlation function \textit{after} the decoherence has become effective. As computed in \cite{Seery_2009_Trispec_GE}, the amplitude of the 4-point correlation function mediated by GE, $\expval{\zeta_{\mathbf{k}_1}\zeta_{\mathbf{k}_2} \zeta_{\mathbf{k}_3} \zeta_{\mathbf{k}_4}}^\text{GE}$, is dependent on the sum over the polarization factors (through the contraction of the momenta $k_i$ of the scalar perturbations with the polarization tensors), i.e. 

\begin{equation}
\expval{\zeta_{\mathbf{k}_1}\zeta_{\mathbf{k}_2} \zeta_{\mathbf{k}_3} \zeta_{\mathbf{k}_4}}^\text{GE} \propto \sum_{P=+,\times}\epsilon_{ij}^P(\mathbf{k}_{12})k_1^i k_2^j\,\epsilon_{lm}^P(\mathbf{k}_{34})  k_3^l k_4^m
\label{eq: sum over polariz}
\end{equation}
where $\mathbf{k}_{12}\equiv \mathbf{k}_1 + \mathbf{k}_2$ (idem for $\mathbf{k}_{34}$).\\
These momenta come from the first order derivatives (in Fourier space) of the scalar perturbations the action \eqref{eq: action for GE}.\\
However, the sum over polarizations in Eq. \eqref{eq: sum over polariz} is not present if the polarization of exchanged graviton is a specific one. This would be the case if, after the interaction between the gravitons and scalar modes around the time of horizon crossing of the gravitons (when they were still in a quantum state), the graviton 2-state decoheres; both gravitons have the same polarization, but now it is just one of the two eigenstates (either $+$ or $\times$, not a superposition) since the state is already classical.

After some treatment, the expression inside the sum in \eqref{eq: sum over polariz} shows dependence on the angles\footnote{The angle $\phi_a$ is the polar angle of the spherical coordinates in an orthonormal basis formed by $\mathbf{k}_{ab}\equiv~\mathbf{k}_a +~\mathbf{k}_b$, and two unitary vectors orthogonal to $\mathbf{k}_{ab}$.} $\phi_a$,  given by the momenta $\mathbf{k}_a$ of the respective inflatons, following the notation from \cite{Seery_2009_Trispec_GE}.

We also note, as shown in \cite{Seery_2009_Trispec_GE}, that in the counter-collinear limit where $|\mathbf{k}_{12}|\ll |\mathbf{k}_1|\approx |\mathbf{k}_2|$, $|\mathbf{k}_3|\approx |\mathbf{k}_4|$, the 4-point correlation through GE has an intuitive physical interpretation; it is the correlation between 2-point functions of scalar perturbations (for example $\expval{\zeta_{{\mathbf{k}_1}}\zeta_{{\mathbf{k}_2}}}$ and $\expval{\zeta_{{\mathbf{k}_3}}\zeta_{{\mathbf{k}_4}}}$) induced by a low energy, gravitational wave of momentum $\mathbf{k_{12}}$ (or graviton):
\begin{equation}
\label{eq: counter collinear lim}
    \lim_{\begin{smallmatrix}
    k_{12}\to 0\\
    k_{34}\to 0
    \end{smallmatrix}}\expval{\zeta_{\mathbf{k}_1}\zeta_{\mathbf{k}_2} \zeta_{\mathbf{k}_3} \zeta_{\mathbf{k}_4}}^\text{GE} = \expval{\expval{\zeta_{\mathbf{k}_1}\zeta_{\mathbf{k}_2}}\expval{ \zeta_{\mathbf{k}_3} \zeta_{\mathbf{k}_4}}}^\text{GE}
\end{equation}
Note that the low momentum $|\mathbf{k}_{12}|$ of the exchanged graviton will set the inverse distance between our laboratories/patches $A$ and $B$ needed to perform a Bell experiment. This gravitational wave/graviton will exit the cosmological horizon during inflation earlier than the inflatons/scalar perturbations (after imprinting the non-local correlations due to entanglement, which occur within the horizon), and could be considered as a fixed classical background into which they exit (see pages 13-15 of \cite{Seery_2009_Trispec_GE}, or \cite{Maldacena:2002} for further discussion).

Let us now construct out Bell-violating observable as described in section \ref{subsec: Bell ineq and experiment} (i.e. an observable that exceeds the classical expectation value):
\begin{equation}
    S^\text{QM}=C(\theta,\phi) + C(\theta',\phi) + C(\theta,\phi') - C(\theta',\phi')
    \label{eq: observable S}
\end{equation}
where
\begin{equation}
    C(\theta,\phi) =  \expval{A(\theta)B(\phi)}\,,
    \label{eq: C=AB}
\end{equation}
and $\expval{\cdot}$ is an average over multiple realizations of the quantum state. $A(\theta)$ and $B(\phi)$ are quantities that depend on the polarization of the corresponding graviton. $A(\theta)$ corresponds to the ``observation'' of the graviton's polarization by Alice's, same goes for $B(\phi)$. Thus, $C(\theta,\phi)$ describes the strong \textbf{correlations} caused by \textbf{entanglement}. The angles $\theta$ and $\phi$ describe the variable that is sensible to the polarization of the graviton during the imprinting process. Thus, they are related to some late-time variable measurable by the final observer. In using the GE (in the counter-collinear limit) as the imprinting process, we take these variables to be angles $\theta \sim \phi_1^A$ and $\phi \sim \phi_3^B$. 
Specifically, $\phi_1^A$ carries polarization information imprinted on inflatons with $\mathbf{k}_1,\,\mathbf{k}_2$, i.e. those contracted with the polarization tensor in Eq. \eqref{eq: sum over polariz} corresponding to the graviton vertex $A$ at Alice's location -- see Fig. \ref{fig: Imprinting}. Same logic applies to $\phi_3^B$ with inflatons with $\mathbf{k}_3,\,\mathbf{k}_4$ at $B$, and the rest at $A'$ and $B'$.\\
We define our operators for Alice's and Bob's patch as: 
\begin{equation}
\label{eq: definitios of A and B operators}
    A(\theta) = \epsilon_{ij}^{P_A}(\phi_1^A) k_1^i k_2^j \,\,\,\,;\,\,\,\,B(\phi) = \epsilon_{lm}^{P_B}(\phi_3^B)k_3^l k_4^m
\end{equation}
(same goes for quantities with ``prime''). Since in our setup the gravitons were entangled and have the same polarization, we write $P_A=~P_B=P_{A'}=P_{B'}\equiv P$.\\

These operators can be written in terms of trigonometric functions (depending on the specific polarization):
\begin{align}
\label{eq: operator as cosine}
    \epsilon_{ij}^P(\mathbf{k}_{ab})  k_a^i k_b^j = \mathcal{N} \cos\left(2\phi_a - \delta_P \right)
\end{align}
where $\delta_+ = 0$ and $\delta_\times = \pi/2$. To get this expression we used the following properties\footnote{The factor $\mathcal{N} = -k_{a}k_{b}\sin\theta_a\sin\theta_b$ will be used as normalization afterwards in order to have the canonical Bell-violating expectation value for the observable \eqref{eq: observable S}. The angles $\theta_{a}$ are the azimuthal angles between vectors $\mathbf{k}_a$ and $\mathbf{k}_{12}$, i.e. $\cos\theta_a = \hat{\mathbf{k}}_a\cdot\hat{\mathbf{k}}_{12}$, see \cite{Seery_2009_Trispec_GE} for further clarification.}:
\begin{equation}
    k_{1,3}\sin\theta_{1,3} = k_{2,4}\sin\theta_{2,4}\,\,\,;\,\,\,\phi_{2,4} = \phi_{1,3} + \pi
\end{equation}

We can construct our $S^\text{QM}$ operator in \eqref{eq: observable S} as:
\begin{align}
    S^\text{QM}  & = \expval{A(\theta)B(\phi)} + \expval{A(\theta')B(\phi)} + \expval{A(\theta)B(\phi')} - \expval{A(\theta')B(\phi')} \notag \\ 
    & = \expval{\epsilon_{ij}^{P}(\phi_1^A) k_1^i k_2^j \,\epsilon_{lm}^{P}(\phi_3^B) k_3^l k_4^m} \notag \\
    & + \expval{\epsilon_{ij}^{P}(\phi_1^{A'}) {k_1^i}' {k_2^j}' \, \epsilon_{lm}^{P}(\phi_3^B) k_3^l k_4^m} \notag \\
    &  + \expval{\epsilon_{ij}^{P}(\phi_1^{A}) k_1^i k_2^j \, \epsilon_{lm}^{P}(\phi_3^{B'}) {k_3^l}' {k_4^m}'} \notag\\
    & - \expval{\epsilon_{ij}^{P}(\phi_1^{A'}) {k_1^i}' {k_2^j}' \, \epsilon_{lm}^{P}(\phi_3^{B'}) {k_3^l}' {k_4^m}'}  \label{eq: S observable with explicit e_ij k_1k_2 forms}
\end{align}
Given the proportionality presented in Eq. \eqref{eq: sum over polariz}, we then have:
\begin{align}
     \hspace{-0.3cm} S^\text{QM}  & \propto \expval{\zeta_{\mathbf{k}_1}^A \zeta_{\mathbf{k}_2}^A \zeta_{\mathbf{k}_3}^B\zeta_{\mathbf{k}_4}^B}^\text{GE} +
    \expval{\zeta_{\mathbf{k}_1}^{A'} \zeta_{\mathbf{k}_2}^{A'} \zeta_{\mathbf{k}_3}^B\zeta_{\mathbf{k}_4}^B}^\text{GE} \notag \\
    & +  \expval{\zeta_{\mathbf{k}_1}^A \zeta_{\mathbf{k}_2}^A \zeta_{\mathbf{k}_3}^{B'} \zeta_{\mathbf{k}_4}^{B'}}^\text{GE} - 
    \expval{\zeta_{\mathbf{k}_1}^{A'} \zeta_{\mathbf{k}_2}^{A'} \zeta_{\mathbf{k}_3}^{B'}\zeta_{\mathbf{k}_4}^{B'}}^\text{GE} \notag \\ & \notag \\
   \hspace{-0.3cm} & = \expval{\expval{\zeta_{\mathbf{k}_1} \zeta_{\mathbf{k}_2}}^{A}\expval{\zeta_{\mathbf{k}_3}\zeta_{\mathbf{k}_4}}^{B}}^\text{GE} \notag \\
    & + \expval{\expval{\zeta_{\mathbf{k}_1} \zeta_{\mathbf{k}_2}}^{A'}\expval{\zeta_{\mathbf{k}_3}\zeta_{\mathbf{k}_4}}^{B}}^\text{GE}  \notag \\
    &  + \expval{\expval{\zeta_{\mathbf{k}_1} \zeta_{\mathbf{k}_2}}^{A}\expval{\zeta_{\mathbf{k}_3}\zeta_{\mathbf{k}_4}}^{B'}}^\text{GE} \notag \\ 
    & - \expval{\expval{\zeta_{\mathbf{k}_1} \zeta_{\mathbf{k}_2}}^{A'}\expval{\zeta_{\mathbf{k}_3}\zeta_{\mathbf{k}_4}}^{B'}}^\text{GE}   \label{eq: S for correlation functions}
\end{align}
were we took the expression \eqref{eq: counter collinear lim} in the counter-collinear limit in the last four lines.\\

In general, we can decompose the polarization states of each graviton in states along and orthogonal to the directions given by $\theta$ (for Alice's location) and $\phi$ (for Bob's location) \cite{Gerry}: 
\begin{equation}
    \begin{aligned}
        & \ket{+}_1 = \cos(2\theta)\ket{\theta}_1 - \sin(2\theta)\ket{\theta^\perp}_1 \\
        & \ket{+}_2 = \cos(2\phi)\ket{\phi}_2 - \sin(2\phi)\ket{\phi^\perp}_2 \\
        & \ket{\times}_1 = \sin(2\theta)\ket{\theta}_1 + \cos(2\theta)\ket{\theta^\perp}_1  \\ 
        & \ket{\times}_2 = \sin(2\phi)\ket{\phi}_2 + \cos(2\phi)\ket{\phi^\perp}_2
    \end{aligned}
\end{equation}
were we defined $\theta^\perp =\theta + \pi/4$ (we get $\ket{+^\perp}=\ket{\times}$ for $\theta=0$, and $ \ket{\times^\perp}=-\ket{+}$ for $\theta=\pi/4$) (idem for $\phi$). If one can define $A(\theta)$ such that $A(\theta)=+1$ if the exchanged graviton is in state $\ket{\theta}$ and $A(\theta)=-1$ if it is in state $\ket{\theta^\perp}$ (idem for $B(\phi)=\pm 1$), then for our Bell state $\ket{\Psi^{-}_\text{Bell}}$ in Eq. \eqref{eq: entangled state of gravitons} we have the following expression for the average defined in Eq. \eqref{eq: C=AB}:
\begin{equation}
    C(\theta,\phi)=\cos\left[4(\theta+\phi)\right]
\end{equation}

We expect that this statistical average can be achieved given the fact that our observable, proportional to the 4-point correlation function, is an average over many observations (cosmology as a laboratory allows for a search in multiple basis/setups until such statistics appear). To obtain a Bell-violating observable, we then take the definitions in \eqref{eq: definitios of A and B operators} and choose $\phi_1^A=2\theta$ and $\phi_3^B=2\phi$ in expression \eqref{eq: operator as cosine} (same for quantities with ``prime''), and normalize
by the factor $\mathcal{N}$:
\begin{equation}
    \Tilde{A}(\theta)=\frac{A(\phi_1^A)}{\mathcal{N}}=\cos\left(4\theta-\delta_P\right)
\end{equation}

Then, for an appropriate choice of angles, e.g. $\phi_1^A=\phi_3^B=0$ and $\phi_1^{A'}=\phi_3^{B'}=\pi/16$, 
expression \eqref{eq: S observable with explicit e_ij k_1k_2 forms} could yield a value of $S^\text{QM}>2=\text{max}\left(S^\text{LHV}\right)$. Thus, the proportional observable coming from the 4-point correlation functions (Eq. \ref{eq: S for correlation functions}) could yield a value that implies non-local correlations coming from the entanglement between our original gravitons. We note that this is not a full falculation. To compute exactly the value of $S$ from \eqref{eq: S for correlation functions}, one would have to make more explicit the proportionality relation in the first line of the equation relative to the S-quantity defined in \eqref{eq: S observable with explicit e_ij k_1k_2 forms}.\\

\begin{figure*}
    \centering
    \includegraphics[width=0.8\textwidth]{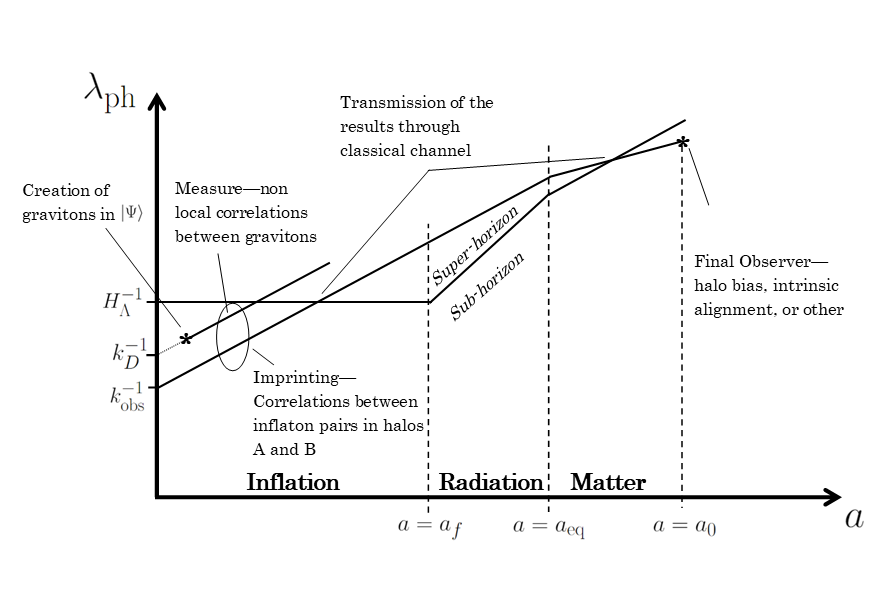}
    \caption{Schematic representation of the whole process described in this work, from the perspective of the evolution of physical scales $\lambda_\text{ph}$ during inflation as a function of the scale factor ($a_f$ denotes the end of inflation, $a_\text{eq}$ the matter-radiation equality, and $a_0$ the scale factor today). $H_\Lambda^{-1}$ denotes the physical Hubble horizon during inflation (constant part) and during the usual Hubble expansion (growing part). $k_D$ corresponds to the tensor modes that will undergo GE with the scalar modes before decoherence (around the time of horizon crossing $k_D|\eta|\sim 1$). $k_\text{obs}$ corresponds to the scalar modes in which we aim to find correlations as a prove of non-local nature (thus quantumness) of inflation. If we are looking at intrinsic alignment between halos, $k_\text{obs}\gtrsim 1$ $\text{Mpc}^{-1}$, so that horizon exit occurs in the last e-folds of inflation, and is observable at the appropriate scale today.}
    \label{fig: time evolution}
\end{figure*}

Let us physically interpret the process of imprinting non-local effects represented in Fig. \ref{fig: Imprinting}. Each instance of GE is represented by one diagram like the one in Fig. \ref{fig: feynmann diag for 4 scalar int}. Since both gravitons are at Alice's and Bob's location, each pair of scalars (located at its respective subhalo) receives a contribution/exchange from both gravitons (which is why we see four diagrams of GE). The two subhalos at each halo A and B correspond to the two different measures that need to be performed in each laboratory. The different measures of the two gravitons at different parts of the two halos will be correlated. The correlation is given by the entanglement between the original gravitons, that latter decohered into classical graviton polarization states that have memory of this correlation, see Sec. \ref{subsec: process - general idea}. Thus, we hint that there is an encoding of the non-local correlations between the two exchanged gravitons' polarizations in the 4-point correlation function described in \cite{Seery_2009_Trispec_GE}. Since we are considering scalar fluctuations that undergo GE before horizon crossing (i.e. $k_i|\eta_\text{GE}|\gtrsim 1$), this signal is frozen and preserved, and it gets classically transmitted to a final observer after the end of inflation ($\eta=\eta_f$) (see Fig. \ref{fig: time evolution}).\\

Finally, we hint possible ways to relate the angles $\phi_1^A$ and $\phi_3^B$ in which the 4-point correlation function of scalar fluctuations  and thus $S^\text{QM}$ depends (see Eqs. \ref{eq: S observable with explicit e_ij k_1k_2 forms}, \ref{eq: S for correlation functions}) to some late-time observable potentially detectable today.\\

As shown in \cite{Bellomo_2018} the GE effect on the trispectrum (4-point correlation) can be manifested as a measurable signal in the halo bias, specifically in the non-Gaussian part of the two- and three-point functions correlating dark matter halos\footnote{The specific relation between the halo bias as a function of the observation angles, and the angles $\phi_1$ and $\phi_3$ is left for a future publication.}. Our final observer would thus be the correlation between the halo bias of halo A and halo B as a function of the angles $\phi_1^A$ and $\phi_1^{A'}$, and $\phi_3^{B}$ and $\phi_3^{B'}$.\\

Another option, as previously mentioned, is a signal in the intrinsic alignment coming from the derivatives on the scalar field fluctuations. These could leave an additional imprint on the two subhalos of each patch with the possibility that we could observe it on the large scale structure through their intrinsic alignment. Note that this intrinsic alignment is obtained by performing the (perturbed Taylor) expansion between one subhalo with the second one in the same patch. In this case, the two derivatives are obtained for the same potential: i.e. a tidal effect between these two subhalos linked to the polarization of the graviton. In this case, the angles $\phi_1^A$, $\phi_1^{A'}$, $\phi_3^B$ and $\phi_3^{B'}$ could be related to the computation of intrinsic alignment under different observational angles. Specifically, a combination of them such that a Bell inequality violation in \eqref{eq: S for correlation functions}. \\

The whole process described in this work is summarized in Fig. \ref{fig: time evolution}.

\section{Conclusions}

We have shown that Bell-type quantum correlations can, in principle, be encoded in cosmological observables within minimal single-field inflation. Polarization-entangled graviton pairs imprint their correlations onto scalar modes through the universal scalar--scalar--tensor interaction, leading to a scalar eight-point function from which a CHSH parameter can be constructed. For appropriate momentum configurations, the resulting correlation function reproduces the expected angular dependence and allows a Bell-inequality violation.

This work isolates the polarization structure of a specific partially disconnected contribution. A complete assessment of the signal requires evaluating the remaining momentum integrals and comparing with other contributions, as well as understanding observational prospects given the high correlator order. Nevertheless, the construction provides a concrete route—using only standard inflationary degrees of freedom—to connect primordial entanglement with Bell-type observables.

\section*{Acknowledgments}

Funding for the work of PT, LS and RJ was partially provided by project PGC2018-098866- B-I00 y
FEDER “Una manera de hacer Europa”, and the “Center of Excellence Maria de Maeztu
2020-2023” award to the ICCUB (CEX2019- 000918-M) funded by MCIN/AEI/10.13039/501100011033
 DB acknowledges partial financial support from the COSMOS
network (www.cosmosnet.it) through the ASI
(Italian Space Agency) Grants 2016-24-H.0, 2016-24-H.1-2018 and
2020-9-HH.

\bibliographystyle{unsrt}

\begin{thebibliography}{10}

\bibitem{Mukhanov:1981xt}
Viatcheslav~F. Mukhanov and G.~V. Chibisov.
\newblock {Quantum Fluctuations and a Nonsingular Universe}.
\newblock {\em JETP Lett.}, 33:532--535, 1981.

\bibitem{Guth:1982ec}
Alan~H. Guth and S.~Y. Pi.
\newblock {Fluctuations in the New Inflationary Universe}.
\newblock {\em Phys. Rev. Lett.}, 49:1110--1113, 1982.

\bibitem{Hawking:1982cz}
S.~W. Hawking.
\newblock {The Development of Irregularities in a Single Bubble Inflationary
  Universe}.
\newblock {\em Phys. Lett. B}, 115:295, 1982.

\bibitem{Bardeen:1983qw}
James~M. Bardeen, Paul~J. Steinhardt, and Michael~S. Turner.
\newblock {Spontaneous Creation of Almost Scale - Free Density Perturbations in
  an Inflationary Universe}.
\newblock {\em Phys. Rev. D}, 28:679, 1983.

\bibitem{Mukhanov_1}
V.~Mukhanov.
\newblock {\em {Physical Foundations of Cosmology}}.
\newblock Cambridge University Press, Oxford, 2005.

\bibitem{Kolv_Turner}
M.~S.~Turner E.~W.~Kolb.
\newblock {\em The Early Universe}.
\newblock Westview Press, 1990.

\bibitem{Campo:2003pa}
David Campo and Renaud Parentani.
\newblock {Space-time correlations in inflationary spectra: A Wave-packet
  analysis}.
\newblock {\em Phys. Rev. D}, 70:105020, 2004.

\bibitem{Campo:2003gb}
David Campo and Renaud Parentani.
\newblock {Space-time correlations within pairs produced during inflation, a
  wave packet analysis}.
\newblock {\em Phys. Rev. D}, 67:103522, 2003.

\bibitem{Campo:2005sv}
David Campo and Renaud Parentani.
\newblock {Inflationary spectra and violations of Bell inequalities}.
\newblock {\em Phys. Rev. D}, 74:025001, 2006.

\bibitem{Campo:2005qn}
David Campo and Renaud Parentani.
\newblock {Quantum correlations in inflationary spectra and violation of bell
  inequalities}.
\newblock {\em Braz. J. Phys.}, 35:1074--1079, 2005.

\bibitem{Campo:2007ar}
David Campo, Jens~C. Niemeyer, and Renaud Parentani.
\newblock {Damped corrections to inflationary spectra from a fluctuating
  cutoff}.
\newblock {\em Phys. Rev. D}, 76:023513, 2007.

\bibitem{Campo:2008ju}
David Campo and Renaud Parentani.
\newblock {Decoherence and entropy of primordial fluctuations. I: Formalism and
  interpretation}.
\newblock {\em Phys. Rev. D}, 78:065044, 2008.

\bibitem{Campo:2008ij}
David Campo and Renaud Parentani.
\newblock {Decoherence and entropy of primordial fluctuations II. The entropy
  budget}.
\newblock {\em Phys. Rev. D}, 78:065045, 2008.

\bibitem{Choudhury:2016cso}
Sayantan Choudhury, Sudhakar Panda, and Rajeev Singh.
\newblock {Bell violation in the Sky}.
\newblock {\em Eur. Phys. J. C}, 77(2):60, 2017.

\bibitem{Choudhury:2016pfr}
Sayantan Choudhury, Sudhakar Panda, and Rajeev Singh.
\newblock {Bell violation in primordial cosmology}.
\newblock {\em Universe}, 3(1):13, 2017.

\bibitem{Kanno:2015ewa}
Sugumi Kanno.
\newblock {A note on initial state entanglement in inflationary cosmology}.
\newblock {\em EPL}, 111(6):60007, 2015.

\bibitem{Martin:2015qta}
Jerome Martin and Vincent Vennin.
\newblock {Quantum Discord of Cosmic Inflation: Can we Show that CMB
  Anisotropies are of Quantum-Mechanical Origin?}
\newblock {\em Phys. Rev. D}, 93(2):023505, 2016.

\bibitem{Kanno:2017teu}
Sugumi Kanno and Jiro Soda.
\newblock {Bell Inequality and Its Application to Cosmology}.
\newblock {\em Galaxies}, 5(4):99, 2017.

\bibitem{Martin:2017zxs}
Jerome Martin and Vincent Vennin.
\newblock {Obstructions to Bell CMB Experiments}.
\newblock {\em Phys. Rev. D}, 96(6):063501, 2017.

\bibitem{Martin:2018zbe}
Jerome Martin and Vincent Vennin.
\newblock {Observational constraints on quantum decoherence during inflation}.
\newblock {\em JCAP}, 05:063, 2018.

\bibitem{Martin:2018lin}
J\'er\^ome Martin and Vincent Vennin.
\newblock {Non Gaussianities from Quantum Decoherence during Inflation}.
\newblock {\em JCAP}, 06:037, 2018.

\bibitem{Kanno_2019}
Sugumi Kanno.
\newblock Nonclassical primordial gravitational waves from the initial
  entangled state.
\newblock {\em Physical Review D}, 100(12), December 2019.

\bibitem{kanno2020polarized}
Sugumi Kanno and Jiro Soda.
\newblock Polarized initial states of primordial gravitational waves, 2020.

\bibitem{Danielson:2021egj}
Daine~L. Danielson, Gautam Satishchandran, and Robert~M. Wald.
\newblock {Gravitationally mediated entanglement: Newtonian field versus
  gravitons}.
\newblock {\em Phys. Rev. D}, 105(8):086001, 2022.

\bibitem{Colas:2022kfu}
Thomas Colas, Julien Grain, and Vincent Vennin.
\newblock {Quantum recoherence in the early universe}.
\newblock {\em EPL}, 142(6):69002, 2023.

\bibitem{Prabhu:2022zcr}
Kartik Prabhu, Gautam Satishchandran, and Robert~M. Wald.
\newblock {Infrared finite scattering theory in quantum field theory and
  quantum gravity}.
\newblock {\em Phys. Rev. D}, 106(6):066005, 2022.

\bibitem{DaddiHammou:2022itk}
Aoumeur Daddi~Hammou and Nicola Bartolo.
\newblock {Cosmic decoherence: primordial power spectra and non-Gaussianities}.
\newblock {\em JCAP}, 04:055, 2023.

\bibitem{Ning_2023}
Sirui Ning, Chon~Man Sou, and Yi~Wang.
\newblock On the decoherence of primordial gravitons.
\newblock {\em Journal of High Energy Physics}, 2023(6), June 2023.

\bibitem{J.Martin_V.Venin}
Jerome Martin and Vincent Vennin.
\newblock Quantum discord of cosmic inflation: Can we show that cmb
  anisotropies are of quantum-mechanical origin?
\newblock {\em Physical Review D}, 93(2), January 2016.

\bibitem{Bellomo_2018}
Nicola Bellomo, Nicola Bartolo, Raul Jimenez, Sabino Matarrese, and Licia
  Verde.
\newblock Measuring the energy scale of inflation with large scale structures.
\newblock {\em Journal of Cosmology and Astroparticle Physics},
  2018(11):043?043, November 2018.

\bibitem{Calzetta:1995ys}
E.~Calzetta and B.~L. Hu.
\newblock {Quantum fluctuations, decoherence of the mean field, and structure
  formation in the early universe}.
\newblock {\em Phys. Rev. D}, 52:6770--6788, 1995.

\bibitem{Lombardo:2005iz}
Fernando~C. Lombardo and Diana Lopez~Nacir.
\newblock {Decoherence during inflation: The Generation of classical
  inhomogeneities}.
\newblock {\em Phys. Rev. D}, 72:063506, 2005.

\bibitem{Martineau:2006ki}
Patrick Martineau.
\newblock {On the decoherence of primordial fluctuations during inflation}.
\newblock {\em Class. Quant. Grav.}, 24:5817--5834, 2007.

\bibitem{Kiefer:2006je}
Claus Kiefer, Ingo Lohmar, David Polarski, and Alexei~A. Starobinsky.
\newblock {Pointer states for primordial fluctuations in inflationary
  cosmology}.
\newblock {\em Class. Quant. Grav.}, 24:1699--1718, 2007.

\bibitem{Kiefer:2008ku}
Claus Kiefer and David Polarski.
\newblock {Why do cosmological perturbations look classical to us?}
\newblock {\em Adv. Sci. Lett.}, 2:164--173, 2009.

\bibitem{Nelson:2016kjm}
Elliot Nelson.
\newblock {Quantum Decoherence During Inflation from Gravitational
  Nonlinearities}.
\newblock {\em JCAP}, 03:022, 2016.

\bibitem{Burgess:2014eoa}
C.~P. Burgess, R.~Holman, G.~Tasinato, and M.~Williams.
\newblock {EFT Beyond the Horizon: Stochastic Inflation and How Primordial
  Quantum Fluctuations Go Classical}.
\newblock {\em JHEP}, 03:090, 2015.

\bibitem{Boyanovsky:2015tba}
D.~Boyanovsky.
\newblock {Effective field theory during inflation: Reduced density matrix and
  its quantum master equation}.
\newblock {\em Phys. Rev. D}, 92(2):023527, 2015.

\bibitem{Burgess_decoherence}
C.P. Burgess, R.~Holman, Greg Kaplanek, Jerome Martin, and Vincent Vennin.
\newblock Minimal decoherence from inflation.
\newblock {\em Journal of Cosmology and Astroparticle Physics}, 2023(07):022,
  July 2023.

\bibitem{Sou:2022nsd}
Chon~Man Sou, Duc~Huy Tran, and Yi~Wang.
\newblock {Decoherence of cosmological perturbations from boundary terms and
  the non-classicality of gravity}.
\newblock {\em JHEP}, 04:092, 2023.

\bibitem{Collins_2016}
Hael Collins and Tereza Vardanyan.
\newblock Entangled scalar and tensor fluctuations during inflation.
\newblock {\em Journal of Cosmology and Astroparticle Physics},
  2016(11):059?059, November 2016.

\bibitem{Gong:2019yyz}
Jinn-Ouk Gong and Min-Seok Seo.
\newblock {Quantum non-linear evolution of inflationary tensor perturbations}.
\newblock {\em JHEP}, 05:021, 2019.

\bibitem{Maldacena_2015}
Juan Maldacena.
\newblock A model with cosmological bell inequalities.
\newblock {\em Fortschritte der Physik}, 64(1):10--23, dec 2015.

\bibitem{Bolis:2016vas}
Nadia Bolis, Andreas Albrecht, and Rich Holman.
\newblock {Modifications to Cosmological Power Spectra from Scalar-Tensor
  Entanglement and their Observational Consequences}.
\newblock {\em JCAP}, 12:011, 2016.
\newblock [Erratum: JCAP 08, E01 (2017)].

\bibitem{Bell}
J.~S. Bell.
\newblock On the einstein podolsky rosen paradox.
\newblock {\em Physics Physique Fizika}, 1:195--200, Nov 1964.

\bibitem{CHSH}
John~F. Clauser, Michael~A. Horne, Abner Shimony, and Richard~A. Holt.
\newblock Proposed experiment to test local hidden-variable theories.
\newblock {\em Phys. Rev. Lett.}, 23:880--884, Oct 1969.

\bibitem{Gerry}
P.~L.~Knight C.~Gerry.
\newblock {\em Introductory Quantum Optics}.
\newblock Cambrige University Press, 2004.

\bibitem{Seery_2009_Trispec_GE}
David Seery, Martin~S Sloth, and Filippo Vernizzi.
\newblock Inflationary trispectrum from graviton exchange.
\newblock {\em Journal of Cosmology and Astroparticle Physics},
  2009(03):018?018, March 2009.

\bibitem{Maldacena:2002}
Juan~Martin Maldacena.
\newblock {Non-Gaussian features of primordial fluctuations in single field
  inflationary models}.
\newblock {\em JHEP}, 05:013, 2003.

\bibitem{Baumann_book_2022}
Daniel Baumann.
\newblock {\em {Cosmology}}.
\newblock Cambridge University Press, 7 2022.

\bibitem{Mukhanov_2}
Mukhanov.
\newblock {\em Introduction to Quantum Effects in Gravity}.
\newblock Cambridge University Press, 2007.

\bibitem{kiefer1998coherence}
Claus Kiefer, Julien Lesgourgues, David Polarski, and Alexei~A Starobinsky.
\newblock The coherence of primordial fluctuations produced during inflation.
\newblock {\em Classical and Quantum Gravity}, 15(10):L67--L72, 1998.

\bibitem{polarski1996semiclassicality}
David Polarski and Alexei~A Starobinsky.
\newblock Semiclassicality and decoherence of cosmological perturbations.
\newblock {\em Classical and Quantum Gravity}, 13(3):377--391, 1996.

\bibitem{xue2012ir}
Wei Xue, Xian Gao, and Robert Brandenberger.
\newblock Ir divergences in inflation and entropy perturbations.
\newblock {\em Journal of Cosmology and Astroparticle Physics},
  2012(06):035--035, 2012.

\end{thebibliography}

\appendix

\section{Perturbations during inflation} 
\label{sec: Perturbations in inflation}

Let us briefly review the basics of quantum fluctuations generated during inflation. While this is textbook material, it is useful as to setup the notation and make the article self-consistent for readers not specialized in the early Universe. 

Any quantum field fluctuates due to the need to satisfy the equations of motion and the Heisenberg uncertainty principle, implemented through quantization of the field \cite{Mukhanov_2}. The simplest models of inflation include 2 fields: a scalar field $\phi$ called the \textit{inflaton} field, and the metric tensor field $g^{\mu\nu}$ that defines the spacetime. Let us first discuss the fluctuations of the inflaton field.

\subsection{Scalar perturbations}

We assume minimal coupling of the massive scalar inflaton field, with mass $m$,  $\phi(\mathbf{x})$ to gravity (in a spatially flat Friedmann universe). The action for the scalar field is \cite{Mukhanov_2}:
\begin{equation}
    \label{eq: action 1}
    I = \frac{1}{2}\int d^4x \sqrt{-g}\,\biggl\{ g^{\mu\nu}\partial_\mu \phi \partial_\nu\phi - V(\phi) \biggr\}\,,
\end{equation}

where $V(\phi) = m \phi^2$. After substituting $g^{\mu \nu}=a^{-2}\eta^{\mu \nu}$ (where $\eta^{\mu\nu}$ refers to the Minkowski metric), $\sqrt{-g}=a^4$, and in terms of the auxiliary field $f(\eta,\mathbf{x})\equiv ~a(\eta)\phi(\mathbf{x})$, (\ref{eq: action 1}) becomes:
\begin{equation}
    \label{eq: action 2}
    \hspace{-0.1cm} I = \frac{1}{2}\int \text{d}^3\mathbf{x}\,\text{d}\eta\,\left\{(f')^2-(\nabla f)^2 + \frac{a''}{a}f^2 - V(\phi)\right\} ,
\end{equation}
where $'$ denotes derivative with respect to conformal time $\eta$, and $\nabla$ is a vector of spatial derivatives. The factor $a''/a$ is time-dependent; it accounts for the interaction of the scalar field with the expanding gravitational background. It implies that the energy of the scalar field is not conserved, which in quantum field theory leads to particle creation \cite{Mukhanov_2}. Since inflation is well approximated by de Sitter spacetime, 
\begin{equation}
    a(\eta)=-\frac{1}{H_\Lambda\eta}\,\,\,,\,\,\,-\infty<\eta<0
\end{equation}
the factor $a''/a$ becomes $2\eta^{-2}$. \\

We can expand the field $f(\eta, \mathbf{x})$ in its Fourier modes:
\begin{equation}
\label{eq: fourier transf scalar}
    f(\eta, \mathbf{x}) =\int\frac{d^3\mathbf{k}}{(2\pi)^{3/2}}\,f_\mathbf{k}(\eta)\,e^{i\mathbf{k}\cdot \mathbf{x}}\,\,.
\end{equation}
Varying the action \eqref{eq: action 2} with respect to $f$ and substituting the mode expansion \eqref{eq: fourier transf scalar}, we get the following differential equation for the scalar field:
\begin{equation}
\label{eq: DE for scalar field}
    f_k''+\left[k^2 + \frac{m^2}{H^2_\Lambda \eta^2}-\frac{2}{\eta^2}\right]f_k=0\,\,.
\end{equation}
The solutions can be written in general as:
\begin{equation}
    \label{eq: scalar Bunch Davis mode functions}
    f_\mathbf{k} = a_\mathbf{k} u_k(\eta) + a_{-\mathbf{k}}^\dagger u_k^*(\eta) \,\,,
\end{equation}
where $a_\mathbf{k}$ and $a_{-\mathbf{k}}^\dagger$ are integration constants, which upon quantization of the field will be promoted to annihilation and creation operators obeying the commutation relations:
\begin{equation}
    \label{eq: scalar comutation relations}
    \left[\Hat{a}_{\mathbf{k}},\,\Hat{a}_{\mathbf{k}'}^\dagger\right] = \delta(\mathbf{k}-\mathbf{k}')\,,\,\,\,\left[\Hat{a}_{\mathbf{k}},\,\Hat{a}_{\mathbf{k}'}\right] = \left[\Hat{a}_{\mathbf{k}}^\dagger,\,\Hat{a}_{\mathbf{k}'}^\dagger\right]=0\,\,.
\end{equation}

$u_k$ and $u_k^*$ are two linearly independent solutions to equation \eqref{eq: DE for scalar field}, which we call mode functions. They are normalized so $u_k {u_{-k}^*} ' - u_k^* u_{-k}'=i$. In the case of de Sitter spacetime , they are given in terms of Bessel functions (see \cite{Mukhanov_2}, pg. 89).\\

We can divide the behavior of a fluctuation of given wave number $k$ in its early and late time asymptotics.

At early times, $k|\eta|\gg 1$, the physical wavelength $\lambda_\text{ph}\sim a\,k^{-1}\simeq \frac{H_\Lambda^{-1}}{k|\eta|}$ is much smaller than the curvature scale of de Sitter $H_\Lambda^{-1}$, and the fluctuation behaves as in flat (Minkowski) spacetime. These are \textit{sub-horizon modes}. We can neglect the $\eta^{-2}$ term in \eqref{eq: DE for scalar field}, and choose the negative-frequency mode to define the minimal excitation of the inflaton field (i.e. vacuum fluctuations) as \cite{Mukhanov_2}:
\begin{equation}
    u_k^{(\text{sub})}\approx \frac{1}{\sqrt{k}}e^{ik\eta}\,\,.
\end{equation}

As spacetime expands, $|\eta|$ decreases (remember $-\infty<\eta<0$), so for a given mode $k$, the physical wavelength is stretched until it becomes of order of the curvature scale $H_\Lambda^{-1}$ at time $\eta=\eta_k$, when $k|\eta|\sim 1$. Fluctuations of mode $k$ then cross the Hubble horizon and start to feel the curvature of spacetime. These are \textit{super-horizon modes}. At asymptotically late times, for $k|\eta|\ll 1$, the term $k^2$ in \eqref{eq: DE for scalar field} can be neglected, and we have solutions:
\begin{equation}
    \label{eq: super horizon modes}
    u_k^{(\text{super})} = A_k|\eta|^{2} + B_k|\eta|^{-1}\,\underset{\eta\rightarrow 0^-} {\longrightarrow} B_k|\eta|^{-1} \,\,,
\end{equation}
where $A_k,B_k \equiv \text{constant}$.\\

The amplitude of the fluctuations of the inflaton field is \cite{Mukhanov_2}:
\begin{equation}
    \label{eq: aplitude of fluctuations}
    \hspace{-0.2cm} \delta_\phi = \frac{1}{2\pi}a^{-1}k^\frac{3}{2}\,|u_k(\eta)| \approx \left\{\begin{array}{lr}
        \frac{1}{2\pi}\frac{k}{a} &;\,\,\lambda_\text{ph}\ll H_\Lambda^{-1} \\ & \\
       \frac{H_\Lambda}{2\pi}k^\frac{3}{2}  &  ;\,\, \lambda_\text{ph}\gg H_\Lambda^{-1}
    \end{array}\right.
\end{equation}
for sub- and super-horizon modes. We see, after horizon crossing at time $\eta_k\sim -k^{-1}$, the amplitude of the fluctuation remains constant in time; it \textit{freezes}. It will remain almost constant until inflation ends, when the physical size of the Hubble horizon will start growing (because of a decelerated expansion) and catch up with the fluctuation. The modes that became super-horizon the latest  (with smallest $k$) will be the first ones to reenter the horizon, and will be the earliest density perturbations of an otherwise homogeneous and isotropic Universe. 
The fluctuations that reentered right after the end of inflation had an effect on the smallest observable scales\footnote{Of course, the non-linear effects of gravity have completely disrupted these small scales at the present time.}, while those which are reentering our horizon now ($t=13.7$ Gyr) affect the largest scales, i.e. the large scale structure of our Universe.

The need of a gracefully exit of the inflationary stage requires that $H_\Lambda$ is not exactly constant, but decreases very slowly. In regard to the fluctuations, this will make those modes which left the horizon earlier have a slightly larger amplitude than the modes which left after (we say the spectrum is red-tilted towards larger scales). Note that we are not considering interactions terms that could modify the quantum-classical transition of the inflaton field \cite{Maldacena_2015,Campo:2007ar}. Indeed due to interaction terms, $k$ modes are not independent anymore and the environment could play an important role, for example \cite{Burgess_decoherence}.

\subsection{Tensor perturbations - gravitons}

We have seen how fluctuations of a massive scalar quantum field can indeed explain the primordial density inhomogeneities in the Universe. Lets now analyze the fluctuations of the other field, the metric tensor $g^{\mu\nu}$. The propagating modes corresponding to the \textit{transverse} and \textit{traceless} tensor fluctuations of the spacetime metric are what we call \textit{gravitons}. They behave as a minimally coupled, massless scalar field with two degrees of freedom (polarizations), and thus can be described by the same formalism used above for the scalar fluctuations (see \cite{Kolv_Turner}, chapter 8.4). This connection between the scalar and tensor sectors can be written
\begin{equation}
    h^{P}_{\mathbf{k}}(\eta) = \sqrt{16\pi G}\,\chi_{\mathbf{k}}(\eta)\,\,\,\,;\,\,\,\,(P=+,\times)
\end{equation}
where $\chi^{+,\times}_\mathbf{k}$ behave as two minimally coupled, real, massless scalar fields. $h_\mathbf{k}^P(\eta)$ are already the Fourier modes introduced in the expansion:
\begin{align}
    \label{eq: fourier transf tensor}
        h_{ij}(\eta, \mathbf{x})=\int\frac{d^3\mathbf{k}}{(2\pi)^{3/2}}\,\sum_{P=+,\times}\,\epsilon_{ij}^P(\mathbf{k})h_\mathbf{k}^P (\eta) \,e^{i\mathbf{k}\cdot\mathbf{x}}\,\,,
\end{align}

where $\epsilon^+ = \Hat{e}_x\otimes\Hat{e}_x - \Hat{e}_y\otimes\Hat{e}_y$ and $\epsilon^\times = \Hat{e}_x\otimes\Hat{e}_y + \Hat{e}_y\otimes\Hat{e}_x$ are the polarization tensors of the two graviton modes. This tensor polarization basis can be  expressed in matrix, by choosing $\Hat{\mathbf{z}}=\Hat{\mathbf{k}}$... as to visualize the matrix for these polarization tensors, form as:
\begin{equation}
    \epsilon^+(\Hat{\mathbf{z}})=\frac{1}{\sqrt{2}}\begin{pmatrix}
        1 & 0 & 0 \\
         0 & -1 & 0 \\
        0 & 0 & 0 
    \end{pmatrix}\,;\,\,\epsilon^\times(\Hat{\mathbf{z}})=\frac{1}{\sqrt{2}}\begin{pmatrix}
        0 & 1 & 0 \\
        1 & 0 & 0 \\
        0 & 0 & 0 
    \end{pmatrix}
\end{equation}

They have the following properties:
\begin{equation}
    \begin{aligned}
       \label{eq: polariz tensor orthogonality} & \epsilon_{ij}^P(\mathbf{k})\epsilon_{ij}^{P'}(\mathbf{k}) = \delta^{PP'} \\
        & \epsilon_{ij}^P(\mathbf{k})\epsilon_{ij}^{P'}(\mathbf{-k}) = s_P\,\delta^{PP'}
    \end{aligned}
\end{equation}

with $s_P=1$ for $P=+$ and $s_P=-1$ for $P=\times$. Note that we can have (at the same time or alternatively) two other different types of polarizations (L and R) that might be interesting for our purposes. In particular, it could create an intrinsic alignment that could be later measured as a signature of entanglement.

Quantization proceeds as in the scalar case; we expand the Fourier modes in their Bunch-Davies mode functions (same as for the scalar case) \cite{Mukhanov_2,Burgess_decoherence,Kanno_2019}: 
\begin{equation}
    \label{eq: Bunch Davies mode func for grav}
    h^{P}_\mathbf{k}(\eta)=\Hat{b}^{P}_\mathbf{k}u_{k}(\eta)+\left(\Hat{b}^{P}_{-\mathbf{k}}\right)^\dagger u_k^*(\eta)
\end{equation}
where $\Hat{b}_\mathbf{k}^P$ is the annihilation operator of a graviton with momentum $\mathbf{k}$ and polarization $P$ (idem for creation), with normalized commutation relations:

\begin{align}
    \left[\Hat{b}^P_{\mathbf{k}},\,\left(\Hat{b}^{P'}_{\mathbf{k}'}\right)^\dagger\right] = \delta(\mathbf{k}-\mathbf{k}')\delta_{PP'}\,\, , \notag \\ \left[\Hat{b}^P_{\mathbf{k}},\,\Hat{b}^{P'}_{\mathbf{k}'}\right] = \left[\left(\Hat{b}^P_{\mathbf{k}}\right)^\dagger,\,\left(\Hat{b}^{P'}_{\mathbf{k}'}\right)^\dagger\right]=0\,\,\,.
    \label{eq: commutation rel for grav}
\end{align}

\section{Two mechanisms for the production of pairs of gravitons entangled in their polarizations} \label{appendix: calculations for graviton pair creation}

\subsection{Two-to-two scattering of inflatons and gravitons}\label{subappendix: 4th order 2-to-2 scattering}

We present the computation of a non-zero probability amplitude for the two-to-two scattering of incoming inflaton perturbations and outgoing tensor perturbations, where the gravitons are in an entangled state of the form \eqref{eq: entangled state of gravitons}.

Consider the following leading order quartic interaction between 2 scalars and 2 tensor modes \cite{xue2012ir}:

\begin{equation}
    S_{\zeta \zeta \gamma \gamma} = -\frac{1}{2} \int d\eta  d^3x \epsilon a^2 \gamma_{il}\gamma_{lj} \partial_i \zeta \partial_j \zeta
    \label{eq: 4th order zeta zeta gamma gamma action}
\end{equation}

If our modes of interest are subhorizon, we can calculate approximately the amplitude of two scalars scattering into two gravitons. The tree level contribution can be easily read off the interaction term and it is of the form

$$V_{s_1s_2}= \frac{1}{2} \epsilon a^2 p_{1i}p_{2j} \epsilon_{il}^{s_1}(\mathbf{k_1})\epsilon_{lj}^{s_2}(\mathbf{k_2})$$

where $\mathbf{p_1}, \mathbf{p_2}$ are the momenta of the scalars, and $\mathbf{k_1}, \mathbf{k_2}$ are the momenta of the gravitons with $s_1$ and $s_2$ being their respective polarizations. The amplitude of getting two entangled gravitons in the following polarization entangled state:

$$\ket{\Psi} = \ket{\mathbf{k_1},\mathbf{k_2}} \otimes \frac{1}{\sqrt2}\bigg(\ket{++} - \ket{\times \times}\bigg)$$

is just obtained from the previous equation by summing over the two polarization states

\begin{equation}
\mathcal{A}_{\Psi}
\;\propto\;
\frac{1}{2}\,\epsilon a^{2}\,p_{1i}p_{2j}
\left[
e^{+}_{il}(\mathbf{k}_{1})e^{+}_{lj}(\mathbf{k}_{2})
-
e^{\times}_{il}(\mathbf{k}_{1})e^{\times}_{lj}(\mathbf{k}_{2})
\right].
\end{equation}

To show that this is nonzero, it is useful to specialize to the case
\begin{equation}
\mathbf{k}_{2}=-\mathbf{k}_{1}\equiv -\mathbf{k}.
\end{equation}

Choose two orthonormal transverse vectors $\mathbf e_{1},\mathbf e_{2}$ such that
\begin{equation}
\mathbf e_{1}\cdot \mathbf e_{1}=1,
\qquad
\mathbf e_{2}\cdot \mathbf e_{2}=1,
\qquad
\mathbf e_{1}\cdot \mathbf e_{2}=0,
\qquad
\mathbf e_{a}\cdot \hat{\mathbf k}=0.
\end{equation}

The linear graviton polarization tensors are
\begin{align}
\epsilon^{+}_{ij}(\mathbf{k})
&=
\frac{1}{\sqrt{2}}
\left(
e_{1i}e_{1j}-e_{2i}e_{2j}
\right),
\\[4pt]
\epsilon^{\times}_{ij}(\mathbf{k})
&=
\frac{1}{\sqrt{2}}
\left(
e_{1i}e_{2j}+e_{2i}e_{1j}
\right).
\end{align}

They satisfy
\begin{equation}
\epsilon^{s}_{ij}=\epsilon^{s}_{ji},
\qquad
\epsilon^{s}_{ii}=0,
\qquad
k_{i}\epsilon^{s}_{ij}=0.
\end{equation}

Now define the transverse projector
\begin{equation}
P_{ij}(\hat{\mathbf k})=\delta_{ij}-\hat{k}_{i}\hat{k}_{j}.
\end{equation}

Then one finds
\begin{align}
\epsilon^{+}_{il}(\mathbf{k})\epsilon^{+}_{lj}(-\mathbf{k})
&=
\frac{1}{2}
\left(
e_{1i}e_{1j}+e_{2i}e_{2j}
\right),\\
\epsilon^{\times}_{il}(\mathbf{k})\epsilon^{\times}_{lj}(-\mathbf{k})
&=
-\frac{1}{2}
\left(
e_{1i}e_{1j}+e_{2i}e_{2j}
\right).
\end{align}

Therefore,
\begin{equation}
\epsilon^{+}_{il}(\mathbf{k})\epsilon^{+}_{lj}(-\mathbf{k})
-
\epsilon^{\times}_{il}(\mathbf{k})\epsilon^{\times}_{lj}(-\mathbf{k})
=
e_{1i}e_{1j}+e_{2i}e_{2j}
=
P_{ij}(\hat{\mathbf k}).
\end{equation}

Hence the amplitude becomes 

\begin{equation}
\mathcal{A}_{\Psi}
\;\propto\;
\frac{1}{2}\,\epsilon a^{2}\,p_{1i}p_{2j}
P_{ij}(\hat{\mathbf k}),
\end{equation}
or equivalently
\begin{equation}
\mathcal{A}_{\Psi}
\;\propto\;
\frac{1}{2}\,\epsilon a^{2}
\left[
\mathbf p_{1}\cdot \mathbf p_{2}
-
(\mathbf p_{1}\cdot \hat{\mathbf k})(\mathbf p_{2}\cdot \hat{\mathbf k})
\right].
\end{equation}

Obviously by momentum conservation we must also have $\mathbf{p_1} = - \mathbf{p_2}$ and in that case the amplitude becomes

\begin{equation}
\mathcal{A}_{\Psi}
\;\propto\;
\frac{1}{2}\,\epsilon a^{2}
\left[
- \mathbf p\cdot \mathbf p
+
(\mathbf p\cdot \hat{\mathbf k})(\mathbf p\cdot \hat{\mathbf k})
\right].
\end{equation}

Therefore, the amplitude is non-zero unless $\mathbf{p}$ is collinear with $\mathbf{k}$.

\subsection{Inflaton decay into a pair of gravitons: $\zeta\rightarrow \gamma\gamma$} \label{subappendix: inflaton decay calculation}

In this section, we present the non-zero probability amplitude for the decay of an inflaton into two gravitons in the final quantum state \eqref{eq: entangled state of gravitons}. We perform the calculation for a generally polarized 2-state, and show in the final expression that it would be non-zero when specified to the desired Bell-like state.

The 3-vertex ($\zeta\gamma\gamma$) term in the action is \cite{Maldacena:2002}:

\begin{equation}
    I_\text{int}^{(3)}=\frac{M_p^3}{8}\int dt \, d^3\mathbf{x}\,a\epsilon_1 \zeta \partial_l\gamma_{ij} \partial_l\gamma_{ij}
\end{equation}

which gives rise to the interaction hamiltonian (after $d\eta=dt/a$):

\begin{equation}
    H_\text{int}(\eta)=-\frac{M_p^3\epsilon_1 a^2}{8} \int d^3\mathbf{x}\,\epsilon_1 \zeta(\eta,\mathbf{x}) \,\otimes\, \partial_l\gamma_{ij}(\eta,\mathbf{x}) \partial_l\gamma_{ij}(\eta,\mathbf{x})
\end{equation}

After changing $v\equiv aM_p\sqrt{2\epsilon_1}\zeta$ and $h_{ij} \equiv (1/2)aM_p\gamma_{ij}$, and taking the scale factor for the approximately de Sitter background $a\approx -(H\eta)^{-1}$:

\begin{equation}
\label{eq: hamiltonian for new fields in position space}
    H_\text{int}^{(3)}\equiv G(\eta)\int d^3\mathbf{x}\, v(\eta,\mathbf{x})\otimes \partial_l h_{ij}(\eta,\mathbf{x}) \partial_l h_{ij}(\eta,\mathbf{x})
\end{equation}

The function $G(\eta)=-\epsilon_1^{-1/2}\left[2\sqrt{2}M_pa(\eta)\right]^{-1}$ is derived in \cite{Burgess_decoherence}.\\

The hamiltonian \eqref{eq: hamiltonian for new fields in position space} in terms of the field Fourier transforms \eqref{eq: fourier transf scalar}, \ref{eq: fourier transf tensor} becomes:

\begin{align}
\label{eq: full H graviton graviton inflaton}
    & H_\text{int}^{(3)}  = \underbrace{G(\eta)\int d^3\mathbf{x}\,\int \frac{d^3\mathbf{p}}{(2\pi)^{3/2}}\,v_\mathbf{p}(\eta)\,e^{i\mathbf{p}\cdot\mathbf{x}}}_{\text{Scalar sector}}\,\otimes \notag \\
    & \otimes \underbrace{\sum_{P=+,\times}\int \frac{d^3\mathbf{k}d^3\mathbf{q}}{(2\pi)^{3}} \,(\mathbf{k}\cdot\mathbf{q})\,\epsilon_{ij}^P(\mathbf{k})\epsilon_{ij}^{P'}(\mathbf{q})h_\mathbf{k }^P (\eta) \,h_\mathbf{q}^{P'} (\eta) e^{i(\mathbf{k}+\mathbf{q})\cdot\mathbf{x}}}_{\text{Tensor sector}}
\end{align}

Let us now compute the matrix element corresponding to the decay of an inflaton into two gravitons, $\zeta\rightarrow \gamma\gamma$:

\begin{widetext}
\begin{align}
    & \left._\text{in}\langle\right.\zeta;\mathbf{p}|\gamma_1,\gamma_2;\,\mathbf{k},\mathbf{q};\,P_1,P_2\rangle_\text{out} \notag \\
    & = \left._{\zeta,\text{in}}\langle\right.0| \left._{\gamma,\text{in}}\langle\right.0|\, a_\mathbf{p} \,\text{exp}\left(-i\int H_I^{(3)} \,d\eta\right)\,\,b_{\mathbf{k},P_1}^\dagger b_{\mathbf{q},P_2}^\dagger | 0\rangle_{\zeta,\text{out}} | 0\rangle_{\gamma,\text{out}} \notag \\ 
    & \sim \left._{\zeta,\text{in}}\langle\right.0| \left._{\gamma,\text{in}}\langle\right.0|\, a_\mathbf{p} \,\Bigl(-i\int H_I^{(3)} \,d\eta + \dots \Bigr)\,\,b_{\mathbf{k},P_1}^\dagger b_{\mathbf{q},P_2}^\dagger | 0\rangle_{\zeta,\text{out}}| 0\rangle_{\gamma,\text{out}} \notag \\ 
    & \sim \left._{\zeta,\text{in}}\langle\right.0| \, a_\mathbf{p} \,(-i) \int d\eta\,  G(\eta)\int d^3\mathbf{x}\,\int \frac{d^3\mathbf{p}'}{(2\pi)^{3/2}}\,v_{\mathbf{p}'}(\eta)\,e^{i\mathbf{p}'\cdot\mathbf{x}}\,| 0\rangle_{\zeta,\text{out}}\, \otimes \notag \\
    & \hspace{1cm } \, \left._{\gamma,\text{in}}\langle\right.0|\,\, \sum_{P,P'=+,\times}\int \frac{d^3\mathbf{m}\, d^3\mathbf{n}}{(2\pi)^{3}} \,(\mathbf{m}\cdot\mathbf{n})\,\epsilon_{ij}^P(\mathbf{m})\epsilon_{ij}^{P'}(\mathbf{n})h_\mathbf{m }^P (\eta) \,h_\mathbf{n}^{P'} (\eta) e^{i(\mathbf{m}+\mathbf{n})\cdot\mathbf{x}} \,   b_{\mathbf{k},P_1}^\dagger b_{\mathbf{q},P_2}^\dagger \ket{0}_{\gamma,\text{out}} \notag \\
    & \sim \left._{\zeta,\text{in}}\langle\right.0| a_\mathbf{p} \,(-i)\int \,d\eta\, G(\eta)\int d^3\mathbf{x}\, \int \frac{d^3\mathbf{p}'}{(2\pi)^{3/2}} e^{i\mathbf{p}'\cdot\mathbf{x}}\,\left( \Hat{a}_{\mathbf{p}'} u_{p'}+\Hat{a}_{-\mathbf{p}'}^\dagger u_{p'}^* \right) \ket{0}_{\zeta,\text{out}} \otimes \notag \\
    & \hspace{1cm} \left._{\gamma,\text{in}}\langle\right.0| \sum_{P,P'}\int\frac{d^3 \mathbf{m}\,d^3 \mathbf{n}}{(2\pi)^3} (\mathbf{m}\cdot\mathbf{n})  \, 
e^{i(\mathbf{m}+\mathbf{n})\cdot\mathbf{x}}\,\epsilon_{ij}^P(\mathbf{m})\epsilon_{ij}^{P'}(\mathbf{n}) \left\{ \Hat{b}^{P}_\mathbf{m}\Hat{b}^{P'}_\mathbf{n} u_m u_n + \Hat{b}^{P}_\mathbf{m}\left(\Hat{b}^{P' }_{-\mathbf{n}}\right)^\dagger u_m u_n^* \notag \right. \\ 
& \hspace{1cm}  \left. + \left(\Hat{b}^{P}_{-\mathbf{m}}\right)^\dagger \Hat{b}^{P' }_{\mathbf{n}} u_m^* u_n + \left(\Hat{b}^{P}_{-\mathbf{m}}\right)^\dagger \left(\Hat{b}^{P'}_{-\mathbf{n}}\right)^\dagger u_m^* u_n^*\right\}   b_{\mathbf{k},P_1}^\dagger b_{\mathbf{q},P_2}^\dagger \ket{0}_{\gamma,\text{out}}
\end{align}
\end{widetext}

where $P_1,P_2$ are the polarization of the created gravitons. In the second line, we have taken the first order approximation of the Dyson expansion, $e^{-i\int d\eta\, H_I}\sim 1-i\int d\eta\, H_I + \mathcal{O}(H_I^2)$, and in the third line we substituted the relevant interaction Hamiltonian to 3 order. Also, in the last equality, we have substituted the Fourier-space functions $v_\mathbf{k}(\eta)$ and $h_\mathbf{k}(\eta)$ by their expansions \eqref{eq: scalar Bunch Davis mode functions} and \eqref{eq: Bunch Davies mode func for grav} in therms of Bunch-Davis mode functions\footnote{Note that the mode functions $u_k$ for the scalar and tensor sectors are the same at sub-horizon scales.}, and expanded the products.\\

Now, applying the commutation rules for the creating and annihilation operators, and using that $a\ket{0} =~\bra{0}a^\dagger =0 $ and $b\ket{0} = \bra{0}b^\dagger=0$:

\begin{widetext}
    \begin{align}
& \sim (-i)\int d\eta\, G(\eta)\int d^3\mathbf{x} \,\int \frac{d^3\textbf{p}'}{(2\pi)^{3/2}}\,\delta(\mathbf{p}'+\mathbf{p}) \, u_{p}^* \cdot\, \sum_{P,P'}\int  \frac{d^3 \mathbf{m}\,d^3 \mathbf{n}}{(2\pi)^3} (\mathbf{m}\cdot\mathbf{n})  \,\epsilon_{ij}^P(\mathbf{m})\epsilon_{ij}^{P'}(\mathbf{n}) \notag \\
& \hspace{1cm} \cdot\Bigl[\delta(\mathbf{m}-\mathbf{k})\delta(\mathbf{n}-\mathbf{q})\delta_{PP_1}\delta_{P'P_2} + \delta(\mathbf{m}-\mathbf{q})\delta(\mathbf{n}-\mathbf{k})\delta_{PP_2}\delta_{P'P_1}\Bigr]\,u_m u_n \, e^{i(\mathbf{p}'+\mathbf{m}+\mathbf{n})\cdot\mathbf{x}} \notag \\
& \sim (-i)\int d\eta\, G(\eta)\int \frac{d^3\mathbf{x}}{(2\pi)^{9/2}}\, (\mathbf{k}\cdot\mathbf{q}) \, \Bigl[\epsilon_{ij}^{P_1}(\mathbf{k})\epsilon_{ij}^{P_2}(\mathbf{q}) + \epsilon_{ij}^{P_2}(\mathbf{k})\epsilon_{ij}^{P_1}(\mathbf{q})\Bigr]\,u_{p}^* u_k u_q \, e^{i(\mathbf{p}+\mathbf{k}+\mathbf{q})\cdot\mathbf{x}} \notag \\
& \sim (-i)\int d\eta\, G(\eta)\, \frac{1}{(2\pi)^{9/2}}\, (\mathbf{k}\cdot\mathbf{q})  \, \Bigl[\epsilon_{ij}^{P_1}(\mathbf{k})\epsilon_{ij}^{P_2}(\mathbf{q}) + \epsilon_{ij}^{P_2}(\mathbf{k})\epsilon_{ij}^{P_1}(\mathbf{q})\Bigr]\,u_{p}^* u_k u_q\, \delta(\mathbf{p}+\mathbf{k}+\mathbf{q})
\end{align}
\end{widetext}

where in the last line we have integrated the spatial integral together with the complex exponential, obtaining the Dirac delta corresponding to the momentum conservation of the decay.

Now, we boost the frame of reference to the CMF of the inflaton $\mathbf{p}\simeq\mathbf{0}$. This is done by assuming  $||\mathbf{p}||\sim~{\cal H}_k \ll ||\mathbf{k}||\sim  1/L$, which is possible together with the symmetry-breaking mechanism suggested in section \ref{subsec: entangled state of gravitons} for including a mass for the scalar perturbation during a specific window of time. At this stage of the calculation, we set $\mathbf{p}\simeq\mathbf{0}$, then the delta function gives $\mathbf{q} \simeq -\mathbf{k}$. Substituting:
\begin{widetext}
\begin{align}
    & \sim (+i)\int d\eta\, G(\eta)\, \frac{1}{(2\pi)^{9/2}}\, k^2  \, \Bigl[\epsilon_{ij}^{P_1}(\mathbf{k})\epsilon_{ij}^{P_2}(-\mathbf{k}) + \epsilon_{ij}^{P_2}(\mathbf{k})\epsilon_{ij}^{P_1}(-\mathbf{k})\Bigr]\,u_{0}^* u_k^2
\end{align}
\end{widetext}

Now, using the orthogonality properties of the polarization tensors \eqref{eq: polariz tensor orthogonality}, and summing over polarizations for a general polarization of the out state, we have:

\begin{align}
    & \sim \sum_{P_1,P_2 = +,\times} i\int d\eta\, G(\eta)\, \frac{1}{(2\pi)^{9/2}}\, k^2  \, \Bigl[s_P\,\delta_{P_1 P_2} + s_P\,\delta_{P_1 P_2}\Bigr]\,v_{0}^* u_k^2 \notag \\
    & \sim \sum_{P_1=+,\times} i\int d\eta\, G(\eta)\, \frac{2}{(2\pi)^{9/2}}\, k^2  \, s_{P_1}\,u_{0}^* u_k^2
\end{align}

Now, computing a non-zero amplitude for the decay of $\zeta$ into a Bell-like pair of gravitons of the form \eqref{eq: entangled state of gravitons} is straightforward:

\begin{align}
&\left._\text{in}\langle\right.\zeta;\mathbf{p}|\gamma_1,\gamma_2;\,\mathbf{k},\mathbf{q};\,\Psi_\text{Bell}\rangle_\text{out} \notag  \\
& = \left._\text{in}\langle\right.\zeta;\mathbf{p}|\gamma_1,\gamma_2;\,\mathbf{k},\mathbf{q};\,+,+\rangle_\text{out} - \left._\text{in}\langle\right.\zeta;\mathbf{p}|\gamma_1,\gamma_2;\,\mathbf{k},\mathbf{q};\,\times,\times\rangle_\text{out} \notag \\
& \sim i\int d\eta\, G(\eta)\, \frac{2}{(2\pi)^{9/2}}\, k^2  \,u_{0}^* u_k^2\, (s_{+} - s_\times) \neq 0
\end{align}
Note the definition of $s_P = 1$ for $P=+$ and $s_P=-1$ for $P=\times$, as previously introduced in \eqref{eq: polariz tensor orthogonality}.

\vspace{10cm}

\end{document}